\documentclass[a4paper,11pt]{article}

\usepackage{jinstpub} 
\usepackage[font=small,labelfont=bf]{caption} 
\usepackage[subrefformat=parens]{subcaption}

\usepackage{multirow}
\usepackage{array}

\title{\boldmath Advances in the Large Area Picosecond Photo-Detector (LAPPD\textsuperscript{TM}): 8" $\times$ 8" MCP-PMT with Capacitively Coupled Readout}

\author[1]{S. Shin,\note{Corresponding author.}}
\author{M. Aviles,}
\author{S. Clarke,}
\author{S. Cwik,}
\author{M. Foley,}
\author{C. Hamel,}
\author{A. Lyashenko,}
\author{D. Mensah,}
\author{M. Minot,}
\author{M. Popecki}
\author{and M. Stochaj}

\affiliation{Incom, Inc.,\\294 Southbridge Rd, Charlton, MA, U.S.A.}

\emailAdd{sshin@incomusa.com}

\abstract{We present advances made in the Large Area Picosecond Photodetector (LAPPD), an 8" $\times$ 8" microchannel plate photomultiplier tube (MCP-PMT), since pilot production was initiated at Incom, Inc. in 2018. The Gen-I LAPPD utilizes a stripline anode for direct charge readout. The novel Gen-II LAPPD employs an internal resistive thin-film which capacitively couples to a customizable external signal readout board, streamlining production and increasing customer flexibility. The Gen-II LAPPD, with an active area of 373 cm$^2$, is capable of high single photoelectron (PE) gain of $\sim$10$^7$, low dark rates ($\sim$1 kHz/cm$^2$), single PE (SPE) timing resolution of $\sim$65 ps, and $\mathcal{O}$(mm) position resolution. Coupled with a UV-grade fused silica window, the LAPPD features a high quantum efficiency (QE) bialkali photocathode of $>$30\% at 365 nm with spectral response down to $\sim$165 nm. The LAPPD is an excellent candidate for electromagnetic calorimeter (ECAL) timing layers, photon-based neutrino detectors, high energy collider experiments, medical imaging systems, and nuclear non-proliferation applications.}


\begin{document}
\maketitle
\flushbottom

\section{Introduction}

Fast picosecond timing, high sensitivity in the UV to visible spectral range, MHz counting rate capability, low dark count rate, good magnetic field tolerance, sub-mm position resolution, and low cost per unit area are necessary requirements for the photosensors that will be employed in future Nuclear Physics (NP) and High Energy Physics (HEP) accelerator experiments like EIC for DIRC and RICH \cite{EIC, RICH} applications and the LHCb ECAL upgrade for HL-LHC \cite{LHCb} respectively, current and future neutrino experiments like the DUNE Near Detector \cite{DUNE}, ANNIE \cite{ANNIE}, WATCHMAN \cite{WATCHMAN}, THEIA \cite{THEIA}, as well as medical applications like Time-of-Flight Positron Emission Tomography (TOF-PET) systems \cite{TOFPET}. Two candidate photosensor technologies that compete for these requirements are Silicon Photo-Multipliers (SiPMs) \cite{SIPM} and Microchannel Plate Photo-Multiplier Tubes (MCP-PMTs) \cite{hamamatsu}. Despite the obvious advantages of the SiPM technology like high Photon Detection Efficiency (PDE), relatively high gain, counting rate capability etc., it suffers from high dark count rates of $\mathcal{O}$(10 kHz/mm$^2$) at room temperature \cite{MPPC}; this can be mitigated by precise thermal cooling that would significantly increase system complexity and cost. Additionally, SiPM dark rates typically increase after exposure to radiation. On the other hand, MCP-PMTs can be operated at room temperature while maintaining low dark count rates ($<$1 kHz/cm$^2$). The active area of some MCP-PMTs can reach up to 53 mm$^2$ \cite{PLANACON} which is significantly larger than that of SiPMs. The cost per unit area for most MCP-PMTs in the market has been relatively high until Incom Inc. introduced the Large Area Picosecond Photo-Detector (LAPPD) -- the most cost-efficient MCP-PMT available to date featuring an active area of 373 cm$^2$.

The LAPPD is based on novel lead-free 203 $\times$ 203 mm$^2$ ALD-GCA-MCPs (MCPs based on Glass Capillary Arrays functionalized by Atomic Layer Deposition) \cite{POPECKI}. Each MCP, available in either 10 $\mu$m or 20 $\mu$m pore size, features an open area ratio (OAR) of up to 74\% with a length to-diameter (L/D) ratio of 60:1. By adjusting our well-defined ALD procedure of applying emissive and resistive layers to GCAs to produce MCPs, the resistance of each MCP can be controlled: ranging from as low as 1 M$\Omega$ to 10s of M$\Omega$.

Signals are read out on either microstrip anodes applied to the inside of the bottom anode plate, or via a capacitively coupled resistive anode. Strip-line readout LAPPDs \cite{LYASHENKO2020} have demonstrated electron gains of $10^7$, low dark noise rates ($<$1 kHz/cm$^2$), single photoelectron (PE) timing resolution about 60 ps, and single PE spatial resolution under 1 mm RMS, and uniform high QE ($\sim$25\%) bi-alkali photocathodes.

In this article we report on the performance of LAPPDs with the capacitively coupled readout. In this configuration, the MCP charge signal is deposited onto a resistive anode and subsequently read out with a customizable PCB board. LAPPD characterization methods and results are presented for gain, dark rates, QE, and spatial and temporal resolutions.

\section{LAPPD Design and Hardware}

\begin{figure}[!htbp]
\centering 
\includegraphics[width=.3\textwidth]{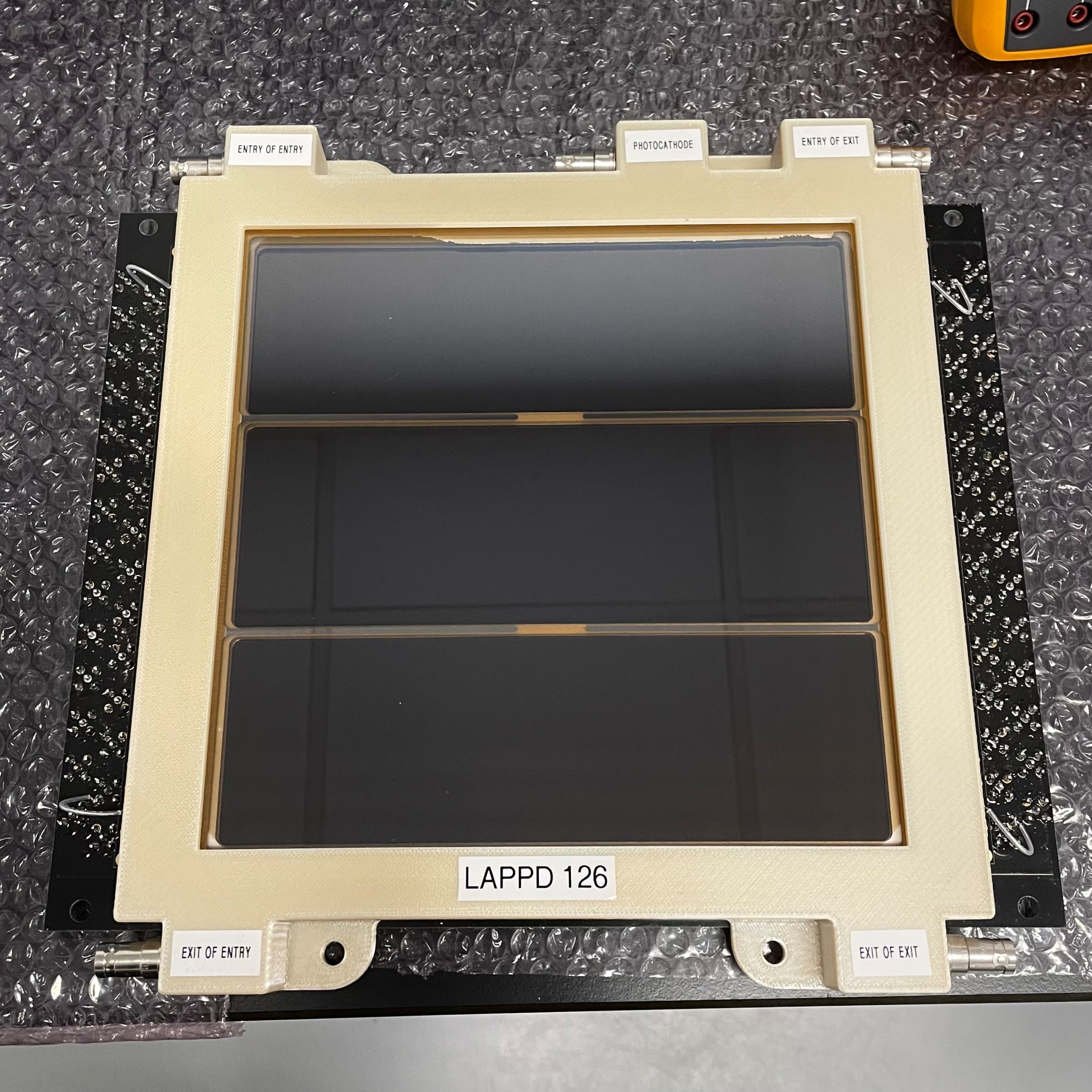}
\quad
\includegraphics[width=.3\textwidth]{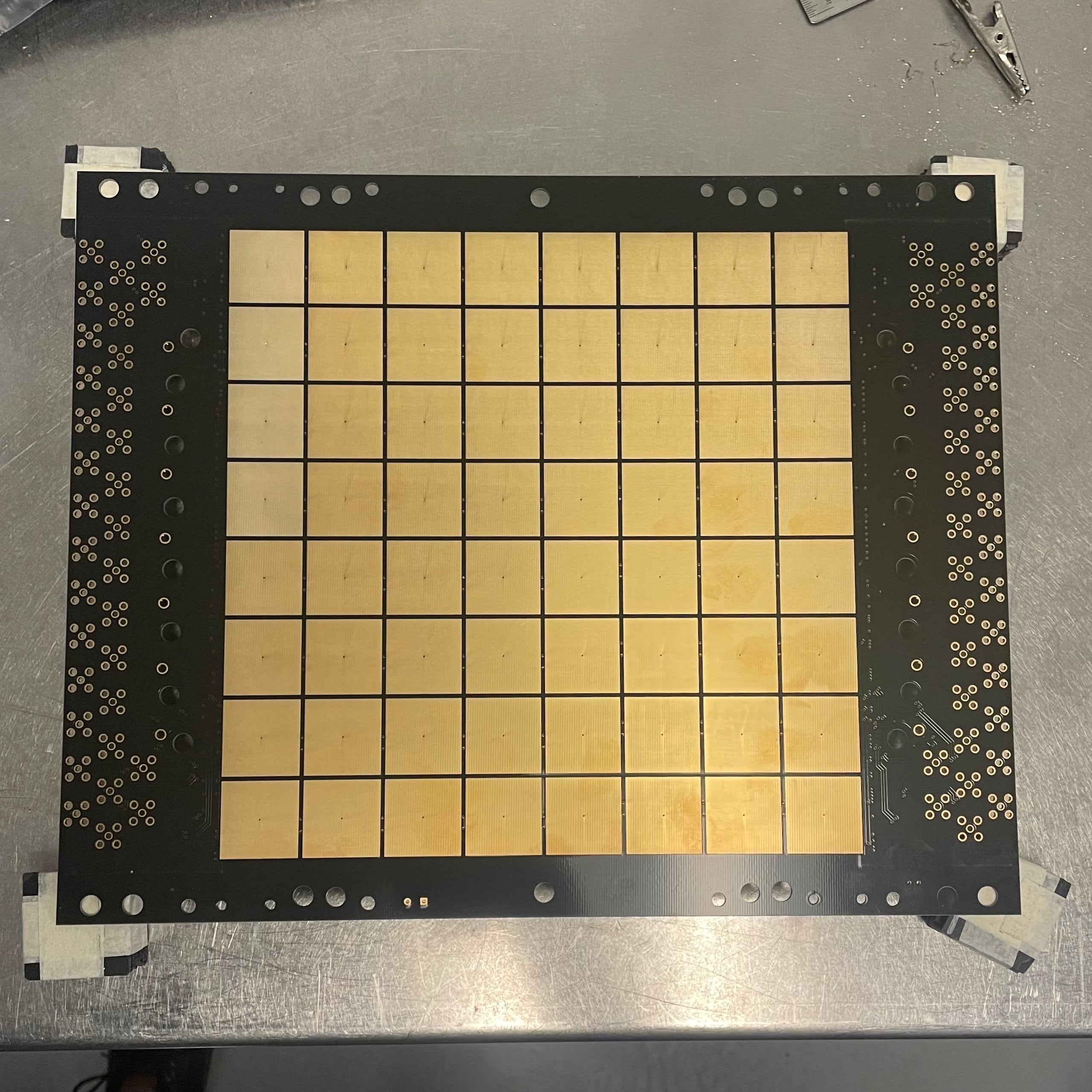}
\caption{ \label{fig:design} LAPPD126 in Ultem housing with independent high voltage connections labeled accordingly (left). Newly designed internal stack components (Rib-spacers) open up the central area of the LAPPD for active detection. 8 $\times$ 8 pixelated readout board with 25 mm pitch pixels that capacitively couple to the internal resistive anode of the LAPPD (right).}
\end{figure}

\begin{figure}[!htbp]
\centering 
\includegraphics[width=.3\textwidth]{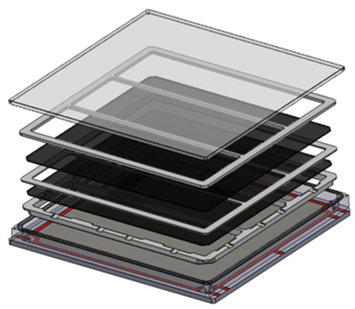}
\quad
\includegraphics[width=.25\textwidth]{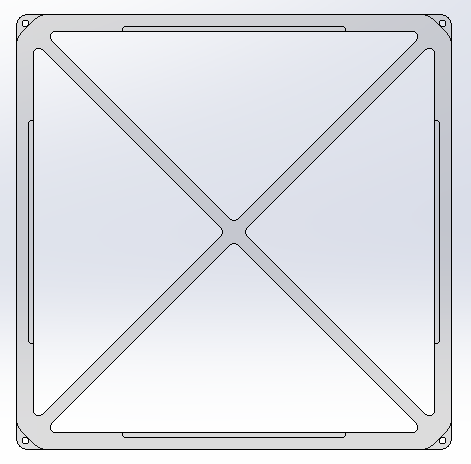}
\caption{ \label{fig:sld} Exploded view of LAPPD and its internal components (left). Previously used X-spacers have been replaced by Rib-spacers (right).}
\end{figure}

\paragraph{General Features.} The LAPPD (Figure \ref{fig:design}) features a chevron pair of independently biased  8" $\times$ 8" ALD-GCA-MCPs enclosed in a hermetically sealed borosilicate glass or ceramic body, internally supported by borosilicate glass spacers. The external dimension of the LAPPD itself is 220 $\times$ 230 mm$^2$, and 300 $\times$ 274 mm$^2$ when fully assembled with the Ultem housing and PCB. High-sensitivity Na$_2$KSb photocathode is deposited onto the fused silica glass entry window of the LAPPD. The substitution of B33 borosilicate glass to fused silica glass for the entry window has extended the spectral response of the LAPPD to the Vacuum Ultraviolet (VUV) range, down to $\sim$165 nm. The internal stack has been redesigned from the previous X-spacer to a rectilinear Rib-spacer, increasing the active area from 350 cm$^2$ to 373 cm$^2$ and opening the central region of the LAPPD to active detection (Figure \ref{fig:sld}). LAPPDs are assembled in an Ultem housing that provides independent high-voltage connections to the photocathode, and the entry and exit faces of each MCP.

\paragraph{Gen-II LAPPD: Capacitively Coupled Readout.}
While the Gen I LAPPD utilizes stripline anode for direct readout, the Gen II LAPPD features an internal resistive film that capacitively couples to a fully customizable external readout board. In the former configuration, the stripline anode is physically coupled to the vacuum packaging of the LAPPD and thus, cannot be modified after sealing. However, with the latter configuration, the decoupling of the photodetector from the readout board permits the use of a customized application-specific pickup pattern. Upon purchase, the Gen II LAPPD is supplied with an 8 $\times$ 8 pixelated readout board (25 $\times$ 25 mm$^2$ pads) (Figure \ref{fig:design}), fully assembled and ready for deployment.

\section{Experimental Methods}

\begin{figure} [!htbp]
\centering
\includegraphics[width=.5\linewidth]{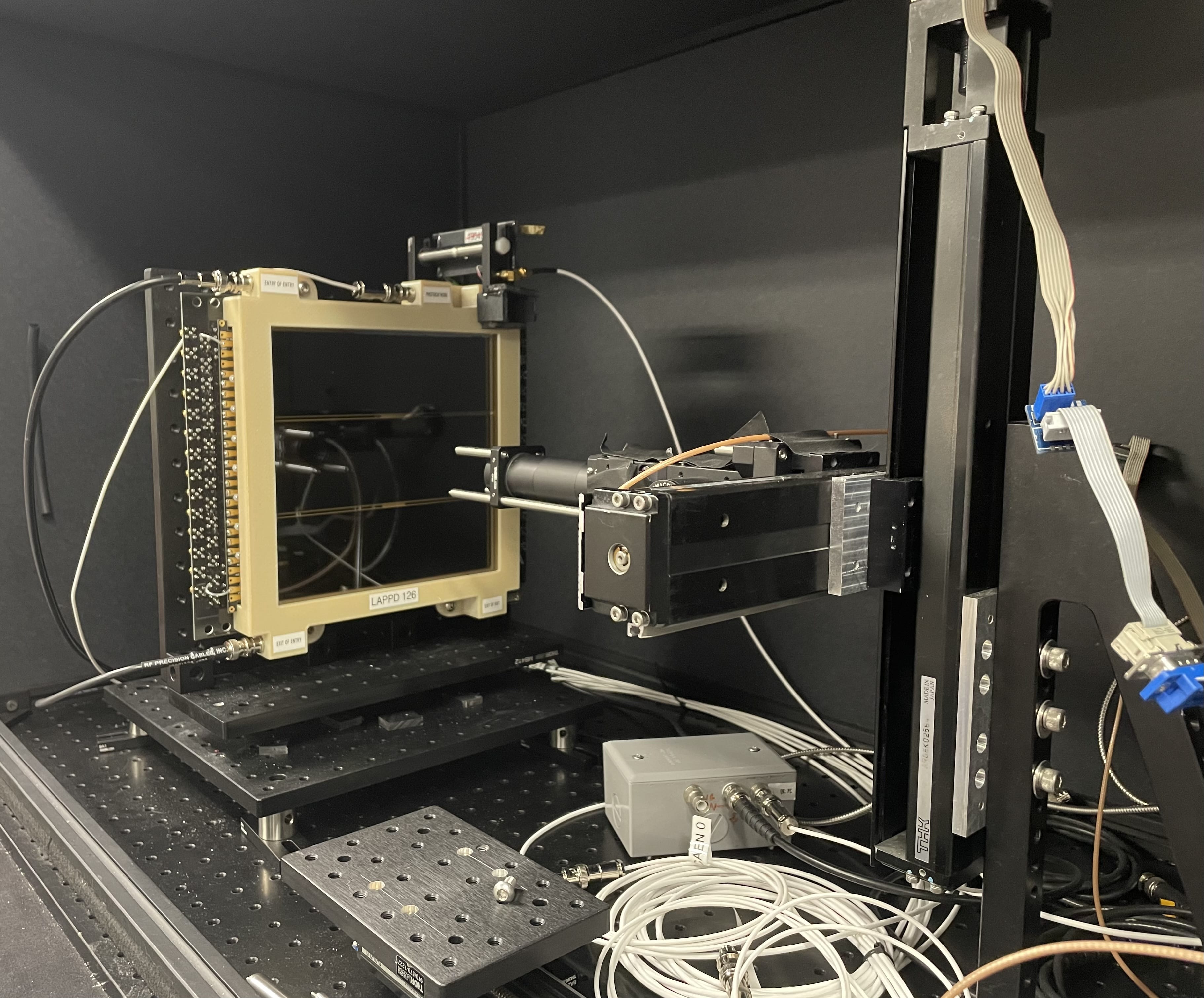}
\caption{LAPPD126 in Ultem housing mounted on and capacitively coupled to a signal readout board on the backplane, ready for testing in the dark box.}
\label{fig:darkbox}
\end{figure}

LAPPD performance tests were performed in a dark box (Figure \ref{fig:darkbox}) with narrow-band LED light sources and signal acquisition hardware. All tests performed and results shown in this paper are those of glass Gen-II LAPPDs with 20 $\mu$m pore size MCPs (65\% OAR), capacitively coupled to an 8 $\times$ 8 pixelated readout board unless explicitly stated otherwise.

\subsection{High QE Photocathode Characterization}

\paragraph{Photocathode Uniformity.}

\begin{figure} [!htbp]
\centering
\includegraphics[width=.7\linewidth]{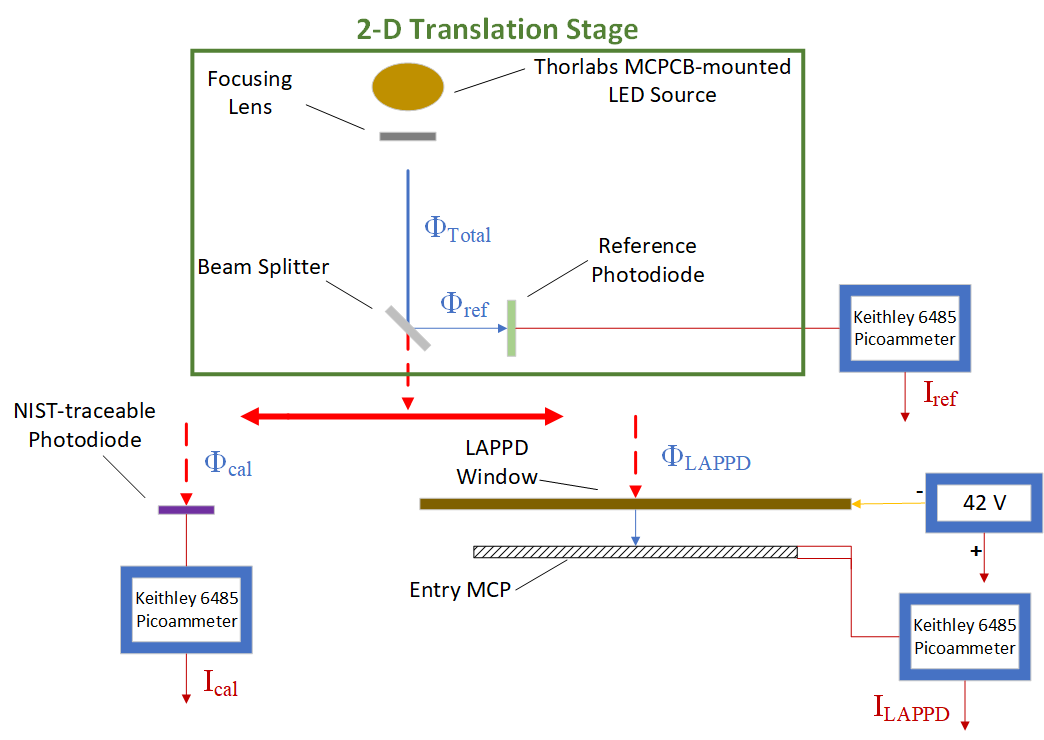}
\caption{QE measurement setup in the dark box shown in Figure \ref{fig:darkbox}. The LED source, beam splitter, and reference photodiode can translate laterally between the NIST-traceable photodiode and the LAPPD window. A similar setup shown in Figure \ref{fig:spectralsetup}, without the 2-D translation stage, is used to test the spectral response of an LAPPD.}
\label{fig:qe}
\end{figure}

The QE measurement setup is structured in such a way that the test stand, which houses the beam splitter, reference photodiode, and the LED beam, is attached to a two-dimensional translation stage that allows movement both to the NIST-traceable photodiode and all along the X-Y plane of the LAPPD window (Figure \ref{fig:qe}). The quantum efficiency map of the photocathode was measured at 365 nm, 420 nm, 455 nm, and 565 nm by scanning the respective narrow-band LED source across the X-Y plane of the LAPPD entry window in 3 mm steps. The LED source was focused by a lens to produce a circular illumination spot of $\sim$2.5 mm diameter on the window. The input light intensity was monitored by a Thorlabs SM1PD2A reference photodiode and measured by a Keithley 6485 picoammeter. With a 42 V bias between the photocathode and the top face of the entry MCP, the photocurrent was recorded with another Keithley 6485 picoammeter. Accounting for the dark current in both, the quantum efficiency at each step was calculated using these two quantities. The reference photodiode was calibrated for every QE scan with a Thorlabs FDS100-CAL NIST-traceable photodiode.

\paragraph{Spectral Response.}

\begin{figure} [!htbp]
\centering
\includegraphics[width=.45\linewidth]{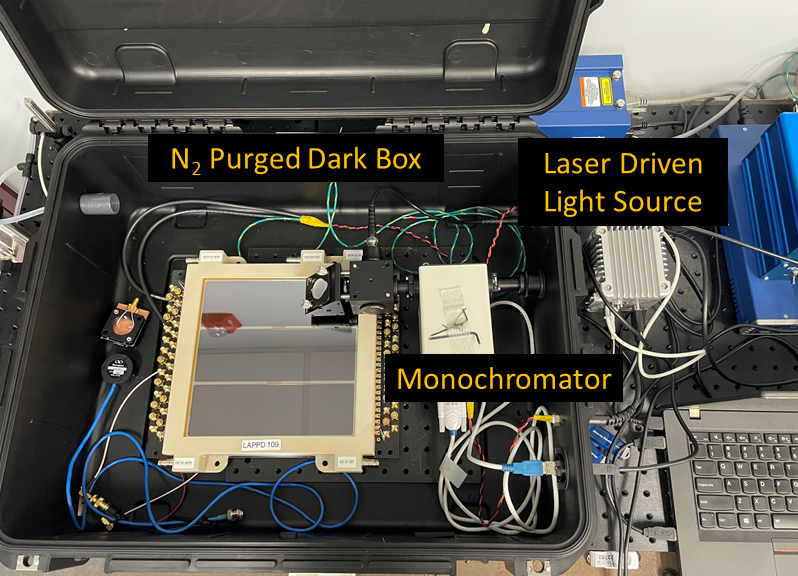}
\caption{A separate benchtop setup is used to measure QE vs wavelength of an LAPPD. The dark box is purged with nitrogen to allow VUV characterization.}
\label{fig:spectralsetup}
\end{figure}

A dedicated nitrogen-purged dark box was built to characterize photocathodes deposited onto fused silica windows in the spectral range: 165 nm to 660 nm (Figure \ref{fig:spectralsetup}). The photocathode QE was measured by irradiating a single point on the active area. The experimental setup is identical to that shown in Figure \ref{fig:qe} without the 2-D translation stage and the substitution of the Thorlabs FDS100-CAL NIST-traceable photodiode with a Opto Diode Corp AXUV100 photodiode. The setup is instrumented with an Energetiq EX-99 Laser Driven Light Source (LDLS) paired with a Digikröm CM110 monochromator.

\paragraph{Quantum Efficiency Derivation.}

The quantum efficiency of the LAPPD photocathode, $\eta_\mathrm{LAPPD}$, is defined as:
\begin{equation} \label{qe}
    \eta_\mathrm{LAPPD} = \frac{I_\mathrm{LAPPD} - I_\mathrm{LAPPD}^\mathrm{dark}}{\Phi_\mathrm{LAPPD} \cdot e}
\end{equation}
where $I_\mathrm{LAPPD}$ and $I_\mathrm{LAPPD}^\mathrm{dark}$ are the current and dark current measured from the top face of the entry MCP respectively, $\Phi_\mathrm{LAPPD}$ is the photon flux (photons per second) incident on the photocathode, and $e$ is the elementary charge. During the QE scan, $\Phi_\mathrm{LAPPD}$ is determined indirectly from the reference photodiode current.

\begin{figure} [!htbp]
\centering
\includegraphics[width=.5\linewidth]{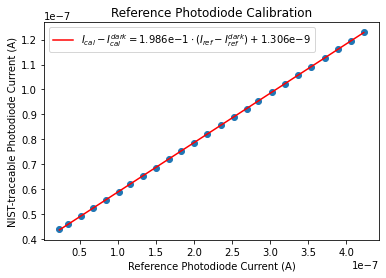}
\caption{An example linear regression analysis between the reference photodiode current and the NIST-traceable photodiode current at 365 nm. This calibration is performed for every QE scan.}
\label{fig:calibration}
\end{figure}

During the calibration of the reference photodiode with the NIST-traceable photodiode, we record currents from both photodiodes, while varying the LED intensity. The following linear relationship is then established:
\begin{equation} \label{fit}
    I_\mathrm{cal} - I_\mathrm{cal}^\mathrm{dark} = m \cdot (I_\mathrm{ref} - I_\mathrm{ref}^\mathrm{dark}) + k,
\end{equation}
where
\begin{itemize}
    \item $I_\mathrm{cal}$ = current measured from NIST-traceable photodiode
    \item $I_\mathrm{cal}^\mathrm{dark}$ = dark current measured from NIST-traceable photodiode
    \item $m$ = slope of the regression line
    \item $I_\mathrm{ref}$ = current measured from reference photodiode
    \item $I_\mathrm{ref}^\mathrm{dark}$ = dark current measured from reference photodiode
    \item $k$ = regression constant
\end{itemize}

We then express the photon flux incident on the NIST-traceable photodiode, $\Phi_\mathrm{cal}$, as:
\begin{equation} \label{NIST flux}
    \Phi_\mathrm{cal} = \frac{m \cdot (I_\mathrm{ref} - I_\mathrm{ref}^\mathrm{dark}) + k}{\eta_\mathrm{cal} \cdot e},
\end{equation}
where $\eta_\mathrm{cal}$ is the quantum efficiency of the NIST-traceable photodiode. $\eta_\mathrm{cal}$ was calculated by multiplying the vendor-provided spectral responsitivity $R$[A/W] by the photon energy $E$[J] = $1.986\mathrm{e}{-16}/\lambda$[nm], where $\lambda$ is the photon wavelength. We assume that the incident photon flux is conserved as the LED target changes from the NIST-traceable photodiode to the LAPPD window, i.e. $\Phi_\mathrm{cal}=\Phi_\mathrm{LAPPD}$. Thus, by substituting Eq. \ref{NIST flux} into Eq. \ref{qe}, we arrive at the final expression:
\begin{equation} \label{finalqe}
    \eta_\mathrm{LAPPD} = \frac{(I_\mathrm{LAPPD} - I_\mathrm{LAPPD}^\mathrm{dark}) \cdot \eta_\mathrm{cal}}{m \cdot (I_\mathrm{ref} - I_\mathrm{ref}^\mathrm{dark}) + k}.
\end{equation}

Typical values of $I_\mathrm{LAPPD}$ and $I_\mathrm{ref}$ are 150 nA and 350 nA respectively.

\subsection{Single Photoelectron Operation}

A PiLas model PiL040-FC pulsed laser with the following specifications was externally triggered to characterize the LAPPD.

\begin{itemize}
    \item \textbf{Wavelength}: 405 nm 
    \item \textbf{Pulse Width}: $\geq$63 ps FWHM (laser intensity dependent)
    \item \textbf{Pulse Frequency}: 0.2 kHz -- 1 MHz
\end{itemize}

In order to operate and characterize the LAPPD in the SPE regime, the laser intensity was attenuated with a Thorlabs NE530B neutral density (ND) filter to a low enough level. The light spot has a diameter of 1 mm on the LAPPD window. We recall the method used in Ref\cite{SPE} to deduce the SPE regime. In order to statistically suppress the probability of producing multiple PEs to less than 10\%, the average number of PEs per laser pulse should be $<$0.21. The laser intensity and ND filters were adjusted such that the LAPPD responded to, on average, 20$\%$ of the laser pulses, hence satisfying this condition.

\subsection{LAPPD Measurements}

\paragraph{Gain.} When an energetic photon strikes the photocathode, an electron is emitted into the vacuum as a PE via the photoelectric effect with a certain probability, i.e. the photocathode's QE. This PE is then amplified as it traverses through both MCPs. This gain was measured as a function of the photocathode voltage, MCP voltage, and event rate using PSI DRS4 \cite{DRS4} waveform samplers. No amplifiers were attached to the front of each DRS4. A 405 nm 63 ps FWHM laser pulse, externally triggered at 3 kHz, was directed at a single point on the LAPPD window in line with the center of a selected pixel of the readout board. The large sheet resistance of the resistive anode (10 k$\Omega$/sq $\sim$ 200 k$\Omega$/sq) allows the electron cloud footprint to be momentarily preserved, generating an image charge on the selected pixel. This ensures a low level of charge sharing with neighboring pixels from the center of one 25 mm pitch pixel.

\begin{figure} [!htbp]
\centering
\includegraphics[width=.5\linewidth]{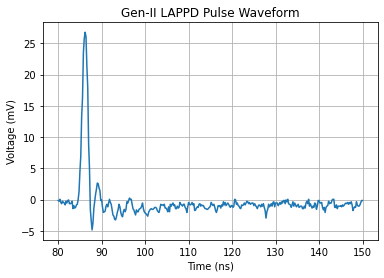}
\caption{Waveform of a typical Gen-II LAPPD pulse. The capacitively detected negative pulse is inverted via an inverting transformer, hence the positive pulse. A capacitive overshoot is observed, followed by a slow return to ground. The DRS4 randomly samples the LAPPD output and whenever the readout of one event is complete, the DRS4 would resume waveform sampling and read out on the next available laser trigger.}
\label{fig:pulse}
\end{figure}

Each pixel is terminated via a 50 $\Omega$ impedance SMA connector and the voltage across this component was sampled at 5 Gs/s. A typical LAPPD signal, $V_{out}(t)$, is shown in Figure \ref{fig:pulse}, with a rise time of $\sim$0.7 ns ($\sim$0.5 ns) for 20 $\mu$m (10 $\mu$m) pore size MCPs; the measured rise time was affected by the analog bandwidth of the DRS4. The charge of each MCP pulse was calculated by integrating the voltage waveform -- scaled down by 50 $\Omega$ at each integration step -- from $t_o$, within a known timeframe of the DRS4 channel, to $t_{zc}$, the zero crossing. The SPE gain is thus the charge divided by the elementary charge, $e$.

\begin{equation} \label{gain}
     G \cdot e = \int_{t_o}^{t_{zc}} \frac{V_{out}(t)}{50 \Omega} dt
\end{equation}

20000 laser pulses were recorded for each test, out of which $<$20\% resulted in LAPPD responses in the form of digitized DRS4 waveforms. The threshold to discriminate the signal pulses from the electronics noise was set to 1.8-2 mV depending on the ambient noise of the DRS4 waveform samplers. Using these data, we plotted a pulse height distribution for each MCP and photocathode voltage configuration. Any waveform with no discernable pulses or any pulses that fall below the given threshold is calculated as "0". When the gain reaches a certain level, the MCPs operate in saturation mode in which further secondary electron emission is limited due to space charge effects inside the channels \cite{hamamatsu}. As a result, a peaked pulse height distribution is observed in the SPE regime. We define the mean value of a peaked pulse height distribution to be the average LAPPD gain.

To measure the gain against event rate, the laser was externally triggered from 0.2 kHz to 400 kHz. With increasing rate, the effective gain decreases as the MCP needs to recharge after each pulse; this "recharge time" is proportional to the MCP's RC time constant \cite{Deadtime}. If a microchannel is struck before it is fully recharged, it will still produce a pulse, but with a lower amplitude. The average gain of the LAPPD was plotted as a function of the observed pulse rate.

\paragraph{Dark Count Rate.} Even in the absence of an external light source, the LAPPD exhibits a constant output of dark current in the form of MCP pulses. The dark count rate is defined as the rate at which these pulses are collected on the anode. Dark rates were measured as a function of the MCP voltage and the photocathode voltage at four separated pixels on the readout board with a 300 MHz bandwidth oscilloscope at a threshold of 4 mV. The same four pixels were used for every dark rate measurements.

The dark count rate was measured at room temperature in two separate configurations. In the first setting, the photocathode was positively biased relative to the entry MCP, thereby suppressing the migration of PEs into the amplification stage. In the second setting, the photocathode voltage was negatively biased relative to the entry MCP, now permitting this migration. Having these two configurations allow for the decoupling of the dominant source of noise, which is typically thermionic emission from the photocathode.

\paragraph{Timing Resolution.} The transit time is defined as the time delay between the initiation of a photoelectron and the arrival of the MCP pulse at the anode. The uncertainty in this time is known as the timing resolution, or transit time spread (TTS), of the LAPPD. The TTS is a crucial factor for timing applications in which discrimination of near-simultaneous events is necessary.

\begin{figure} [!htbp]
\centering
\includegraphics[width=.65\linewidth]{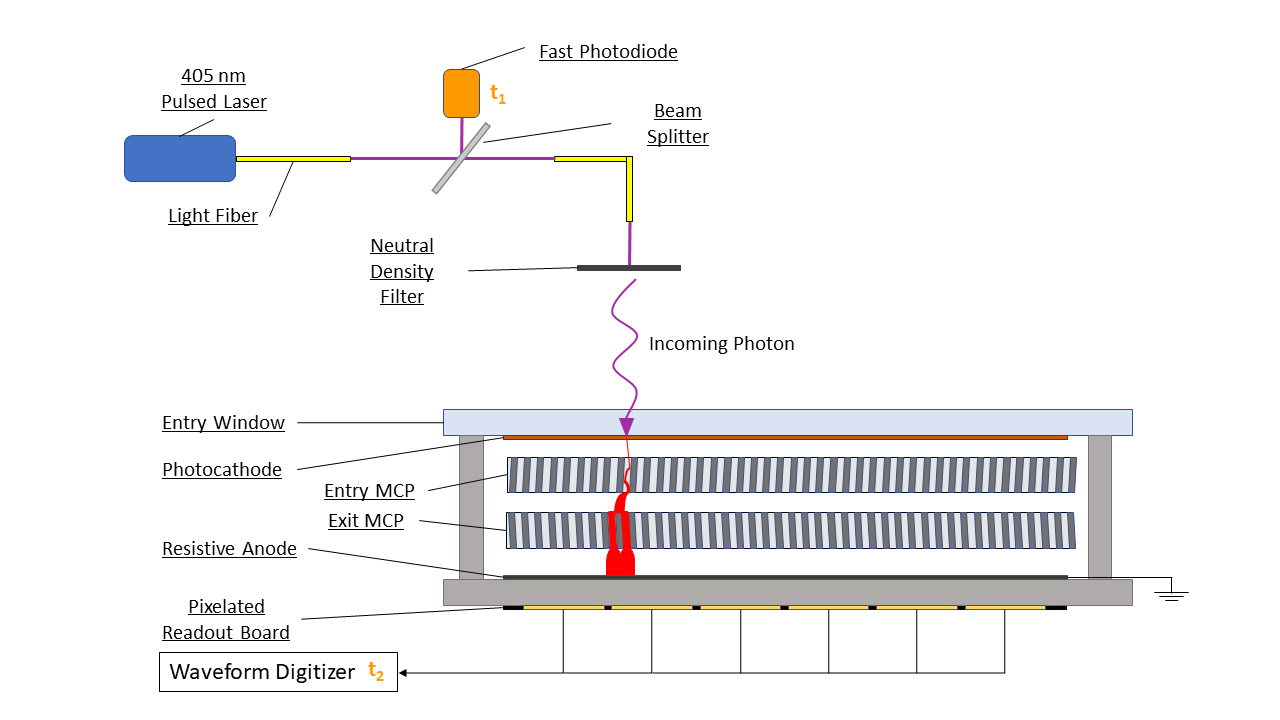}
\caption{Schematic of LAPPD characterization setup in the dark box for gain, timing, and position measurements. $t_1$ and $t_2$ are defined as the time of laser trigger at the fast photodiode and the readout of a digitized pulse respectively. The difference of these two values, $t_2 - t_1$, is measured by using a fast photodiode to directly monitor the 405 nm laser pulse.}
\label{fig:tts}
\end{figure}

A 63 ps FWHM 405 nm laser pulse was directed at a beam splitter, and subsequently onto a Thorlabs DET025AFC fast photodiode and a ND filter, through which the laser intensity is attenuated to the single photon level (Figure \ref{fig:tts}). Therefore, the laser photon that produces a SPE in the LAPPD may arrive at any time within the 63 ps window. At the laser intensity in this example, the laser output has a Gaussian-like profile of light emission vs. time. The 63 ps FWHM corresponds to a $\sigma_\mathrm{LW}$ sigma of 26.8 ps. 

\begin{figure} [!htbp]
\centering
\includegraphics[width=.55\linewidth]{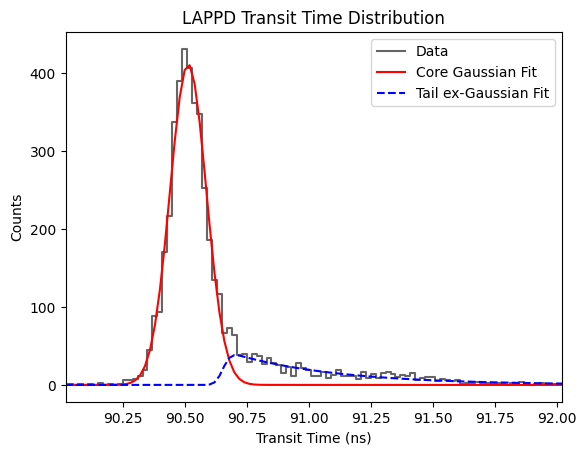}
\caption{Transit time distribution of an LAPPD. The core (red solid line) describes the primary contribution to the distribution. The tail (blue dashed line) represents the late arrival times of both backscattered PEs and secondary electrons emitted from the thin film electrode on the top face of the entry MCP.}
\label{fig:ttd}
\end{figure}

We derive the timing for SPE from the centroid of the MCP pulse. The centroid of an MCP pulse is defined as the weighted mean of the corresponding waveform. The transit time distribution of an LAPPD is then evaluated as the sum of two components: the "core" distribution and the "tail" distribution (Figure \ref{fig:ttd}). A Gaussian fit is used to model the core distribution. An exponentially modified Gaussian (ex-Gaussian) distribution is then fitted to the residual tail distribution. Variations in the core distribution may come from phenomena such as the depth to which the PE advances into the microchannel before striking the walls of the channel. Depending on the OAR of the entry MCP, a certain fraction of PEs will interact with the top interstitial layer -- a thin film electrode -- of the entry MCP. Specifically, PEs may either elastically scatter off the electrode as backscattered PEs or inelastically collide with it and potentially yield secondary electrons. The resulting backscattered PEs or secondary electrons may then enter a microchannel, producing late arrival pulses described by the tail distribution. In this manuscript, we provide results of the core TTS $\sigma_\mathrm{core}$ as a function of the photocathode bias voltage.

Additional variations arise from electronic noise superimposed on the pixel measurement and on the photodiode waveform. There is also a $\sim$$\pm$25-50 ps jitter in the width of the DRS4 timesteps \cite{calibration}, which is nominally 200 ps but not corrected here.

The transit time spread of the core distribution was obtained from the following formula:

\begin{equation} \label{eq7}
     \sigma_\mathrm{core} = \sqrt{\sigma_\mathrm{system}^2 - \sigma_\mathrm{LW}^2},
\end{equation}

where

\begin{itemize}
    \item $\sigma_\mathrm{core}$ = LAPPD core time resolution
    \item $\sigma_\mathrm{system}$ = System time resolution (i.e. sigma of the core Gaussian fit)
    \item $\sigma_\mathrm{LW}$ = laser pulse width (typically increases with laser intensity)
\end{itemize}

\section{Advances in Photocathode Development}

\begin{figure}[!htbp]
\centering 
\includegraphics[width=.3\textwidth]{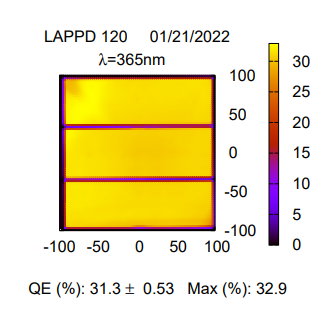}
\qquad
\includegraphics[width=.3\textwidth]{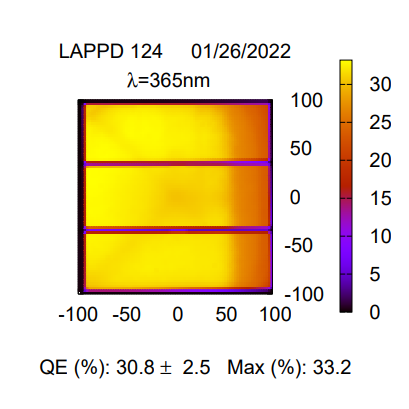}
\qquad
\includegraphics[width=.3\textwidth]{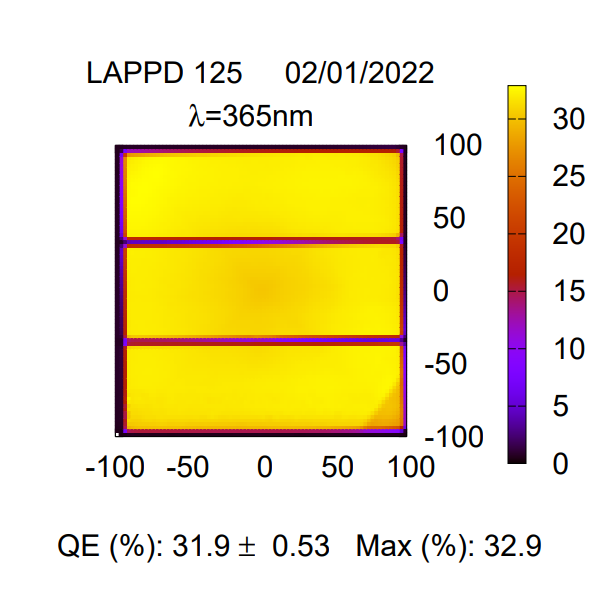}
\qquad
\includegraphics[width=.3\textwidth]{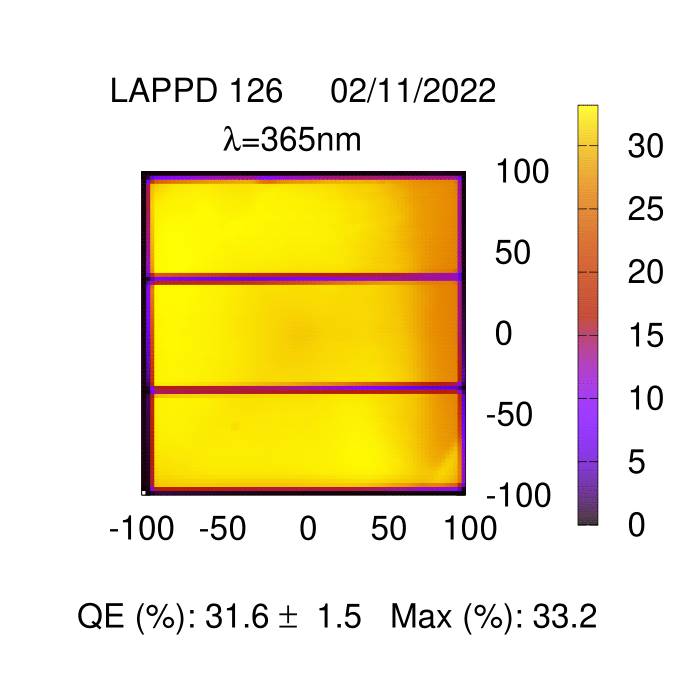}
\caption{\label{fig:365QE} 365 nm QE maps of four consecutively sealed Gen-II LAPPDs. Average QE values and corresponding standard deviations are indicated in each subfigure. Both axes represent the position in mm. Due to internal reflection within the window, the light spot at the photocathode is expected to have a larger diameter than 2.5 mm, with diminishing intensity beyond that. As the width of the rib-spacer is 3.2 mm, the partially blocked light results in a non-zero QE on the rib-spacer.}
\end{figure}

Photon detection efficiency (PDE) is one of the most important characteristics of a photosensor. In PMTs, PDE is defined as the product of the photocathode QE and the photoelectron collection efficiency (CE). The CE is the probability that the PE emitted from the photocathode hits on the first dynode or the microchannel, initiating an electron avalanche and subsequently resulting in an output pulse from the anode. The latter is usually 60-90\% depending on the type of MCPs and operating voltage for MCP-PMTs, making QE a driving factor for achieving a high PDE. The photocathode growth process is largely experimental, requiring active presence and attention. Several factors that govern the performance of a photocathode are optimal temperature control, quality of the vacuum condition, surface treatment of the substrate, and the skills of the workers themselves \cite{UQE}.

\begin{figure}[!htbp]
\centering 
\includegraphics[width=.3\textwidth]{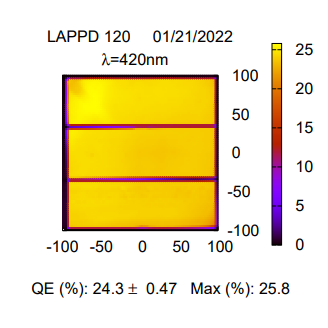}
\quad
\includegraphics[width=.3\textwidth]{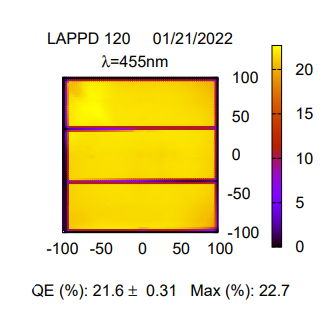}
\quad
\includegraphics[width=.3\textwidth]{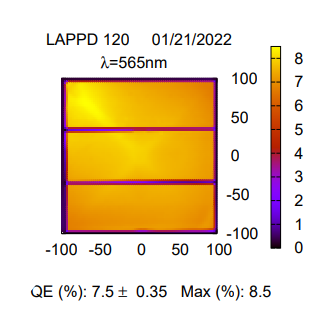}
\caption{\label{fig:fullQE} QE maps of LAPPD120 measured at 420 nm, 455 nm, and 565 nm.}
\end{figure}

A reliable, highly repeatable process for deposition of high QE Na$_2$KSb photocathodes was developed as evidenced by the four consecutively sealed LAPPD120, LAPPD124, LAPPD125, and LAPPD126. QE scans of LAPPD120, LAPPD124, LAPPD125, and LAPPD126 measured at 365 nm are shown in Figure \ref{fig:365QE}. All four tiles demonstrated average QE values exceeding 30\%, with LAPPD120 and LAPPD125 featuring spatial uniformity of $>$98\%; the QE spatial uniformity is defined as the difference between the mean QE value and the standard deviation over the mean QE value. LAPPD124 and LAPPD126 share the same darker QE region which we attribute to sub-optimal variations in temperature distribution during the deposition process due to a faulty heating element. Additional QE scans of LAPPD120 measured at 420 nm, 455 nm, and 565 nm are shown in Figure \ref{fig:fullQE}.

\begin{figure}[!htbp]
\centering 
\includegraphics[width=.3\textwidth]{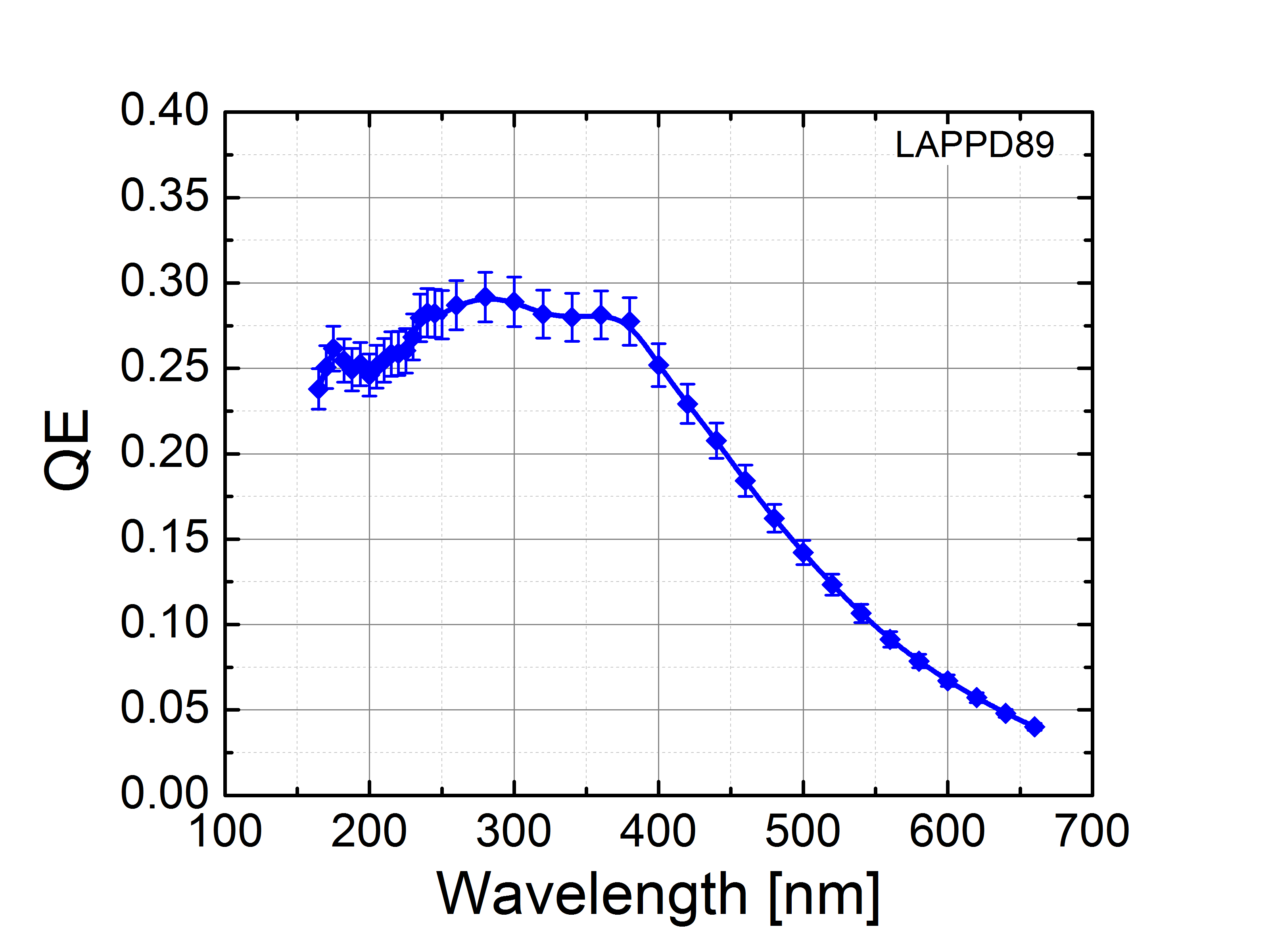}
\quad
\includegraphics[width=.3\textwidth]{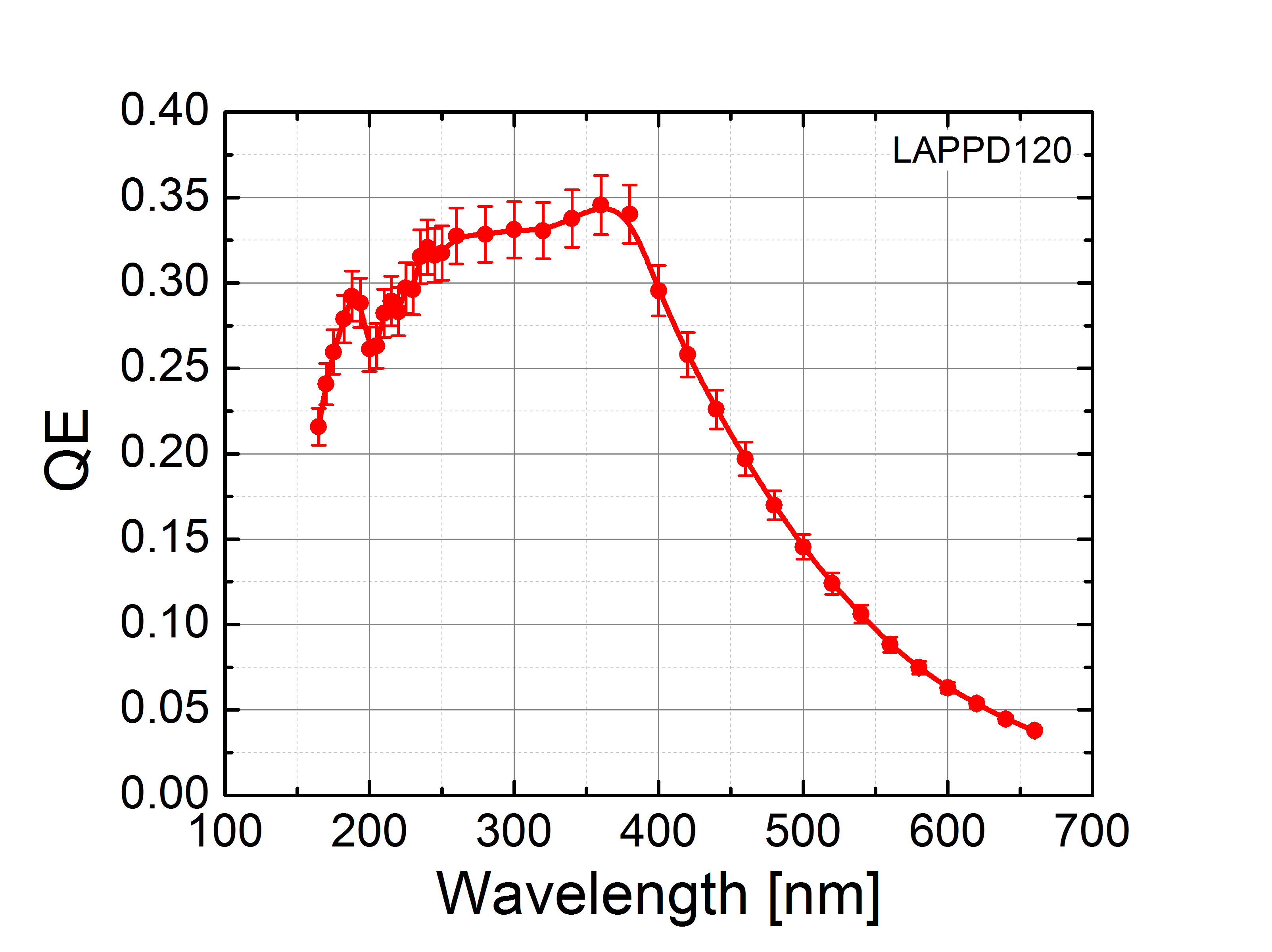}
\quad
\includegraphics[width=.3\textwidth]{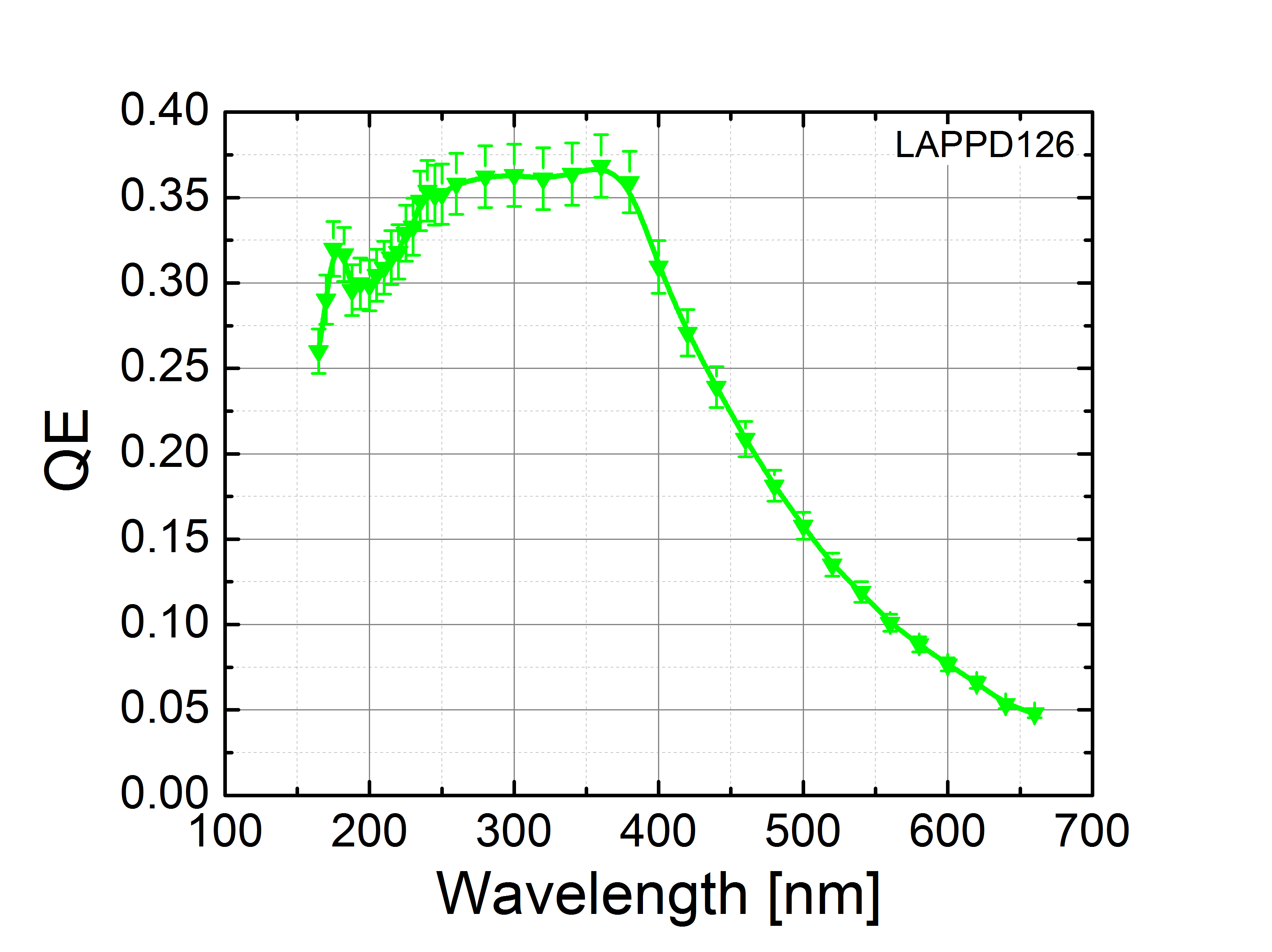}
\caption{\label{fig:VUVQE} QE as a function of wavelength measured for LAPPD89, LAPPD120, and LAPPD126.}
\end{figure}

QE spectra of LAPPD89, LAPPD120 and LAPPD126 with fused silica windows measured in a spectral range between 160 nm and 660 nm are presented in Figure \ref{fig:VUVQE}. The QE spectra of the three tested LAPPDs reveal similarity in their spectral response. As shown in Figure \ref{fig:VUVQE}, Na$_2$KSb photocathode deposited on a fused silica window has a local maximum between 170 nm and 180 nm. There is an elevated plateau between 245 nm and 380 nm followed by a gradual decline. The highest QE value achieved was $\sim36\%$ in LAPPD126 at 360 nm. The systematic error in these measurements was estimated to be about 10\% which represents the error in the spectral responsitivity of the photodiode that was not NIST-calibrated in this setup. This explains the slight discrepancy in the QE measurements for LAPPD126 in Figures \ref{fig:365QE} and \ref{fig:VUVQE}.

\section{Performance Results}

The applied voltage between the bottom face of the entry MCP and the top face of the bottom MCP and that between the bottom face of the exit MCP and the resistive anode were kept at 200 V throughout all measurements.

\subsection{Gain}

\begin{figure}[!htbp]
\centering 
\includegraphics[width=.49\textwidth]{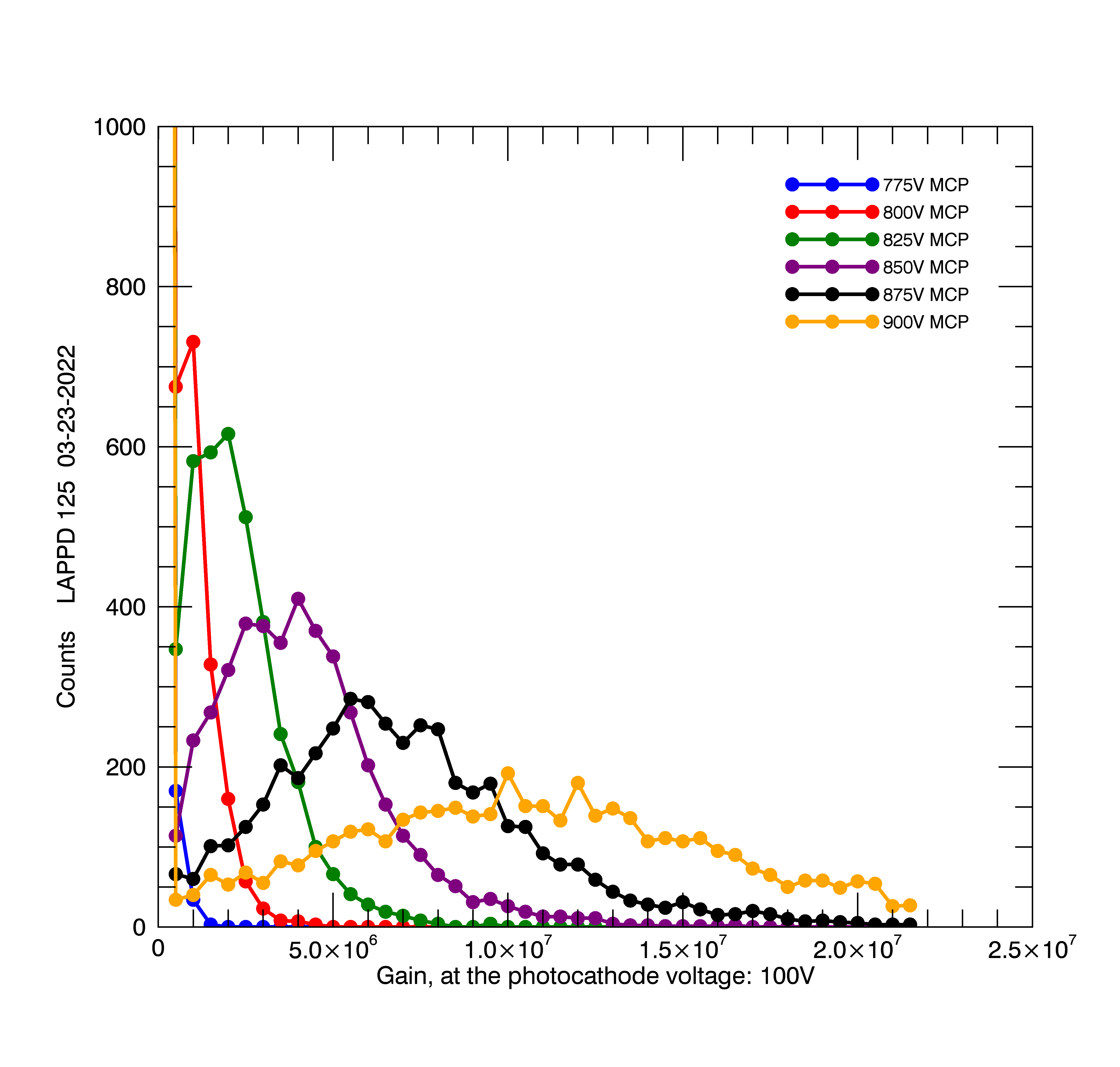}
\includegraphics[width=.45\textwidth]{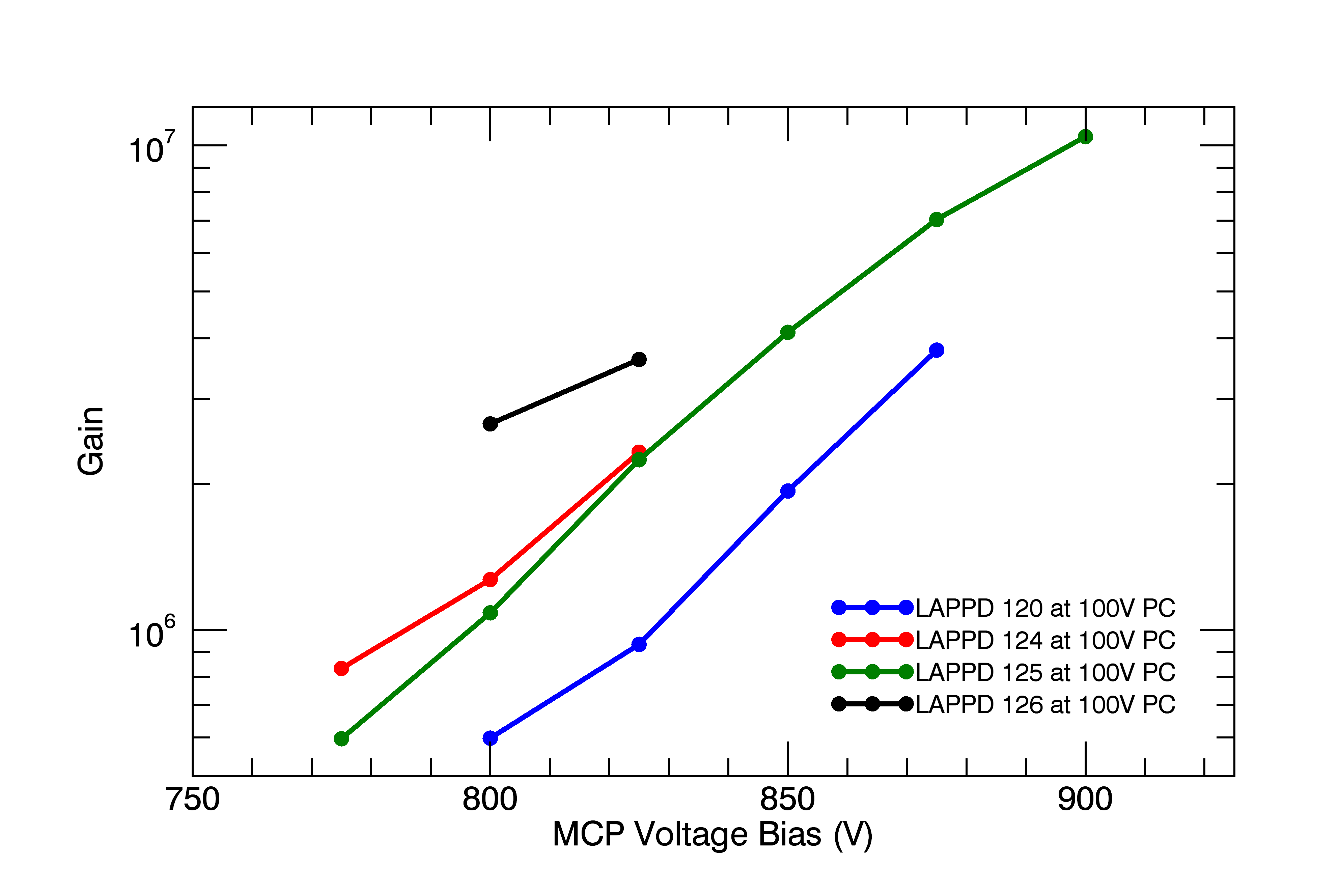}
\caption{Typical peaked pulse height distributions across different MCP voltages (left). Gain as a function of MCP voltage for four different LAPPDs (right). Photocathode bias voltages were fixed at 100 V.}
\label{fig:gain:mcp}
\end{figure}

\begin{figure}[!htbp]
\centering 
\includegraphics[width=.4\textwidth]{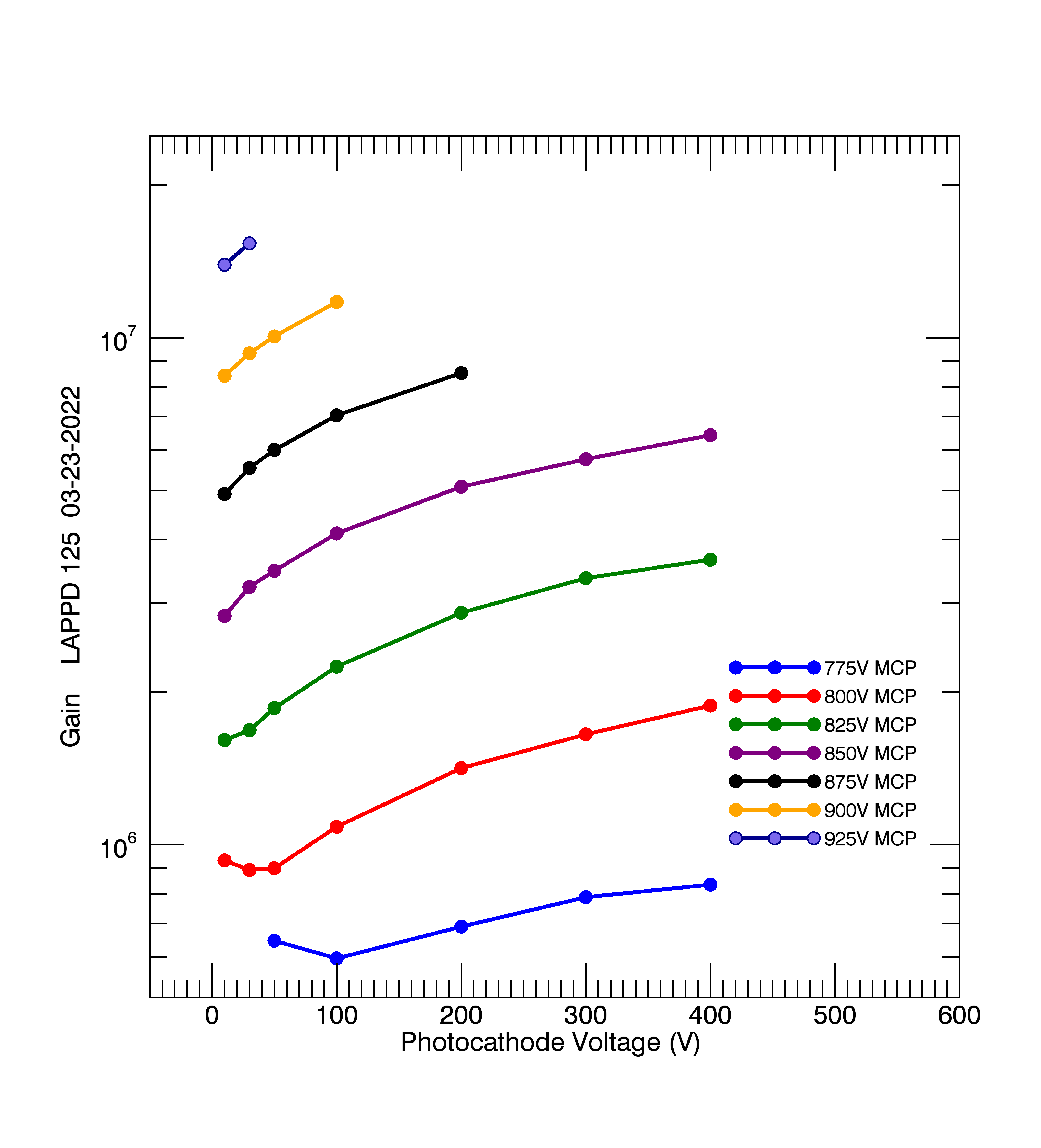}
\includegraphics[width=.4\textwidth]{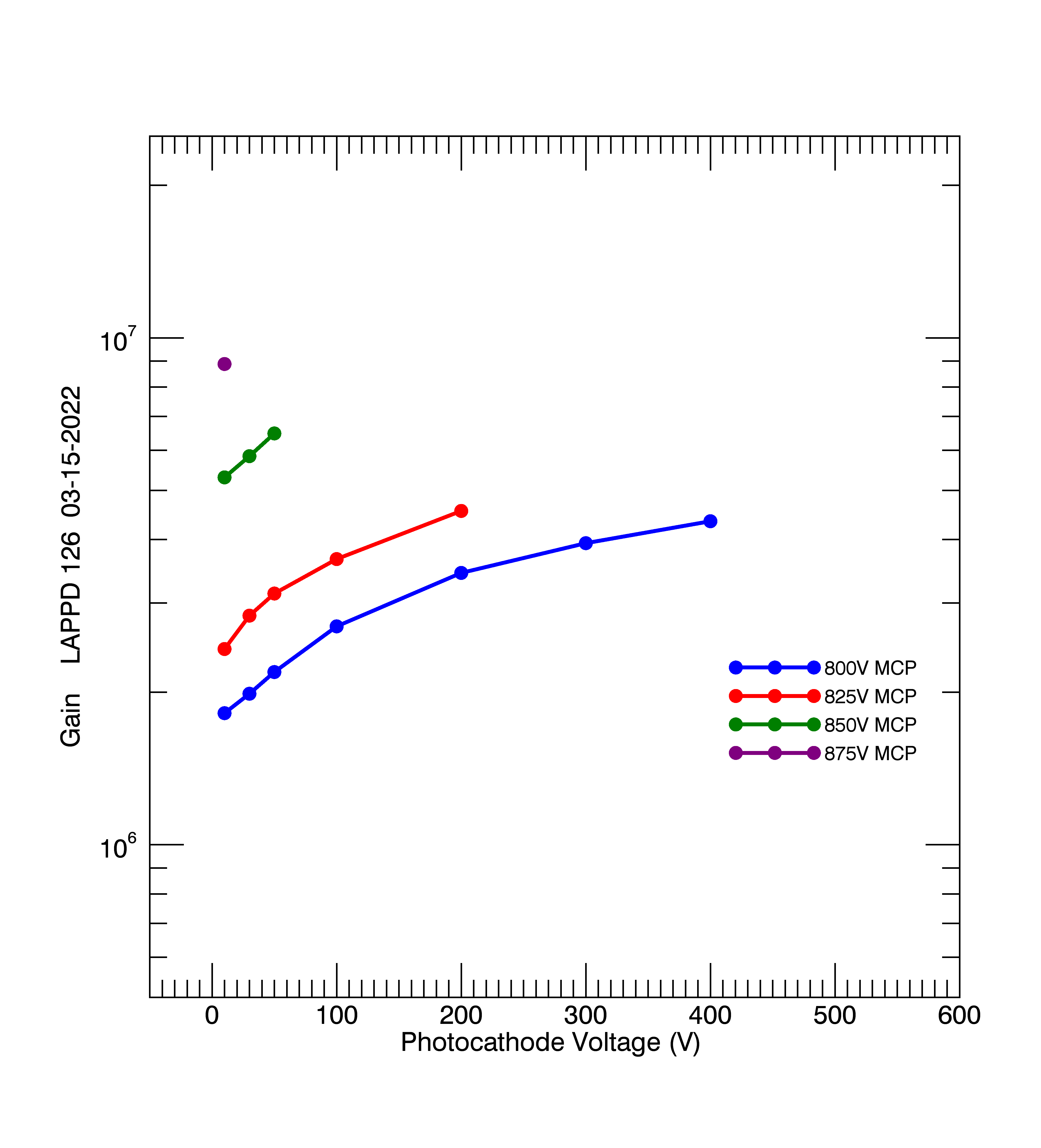}
\caption{Gain as a function of photocathode voltage across different MCP voltages for LAPPD125 (left) and LAPPD126 (right).}
\label{fig:gain:pc}
\end{figure}

All four Gen-II LAPPDs attained gain in the low to mid-10$^6$ range at $>$825 V per MCP at a fixed photocathode bias voltage of 100 V (Figure \ref{fig:gain:mcp}). LAPPD125 demonstrated high gains of $\sim$10$^7$ at 900 V per MCP. Data was taken for different MCP voltages but the measurement for each LAPPD ended when the dark count rate escalated, leading to a state in which the output current was no longer stable.

By increasing the photocathode bias voltage, the LAPPD gain increased but to a lesser extent than in Figure \ref{fig:gain:mcp} as the photocathode voltage only affects the initial acceleration and impact energy of the PE. At 825 V per MCP and a 100 V photocathode bias, LAPPD125 and LAPPD126 demonstrated gains of $2\cdot10^6$ and $3.5\cdot10^6$ respectively (Figure \ref{fig:gain:pc}). However, as the gain at low MCP voltages is likely overestimated as the detection of pulses decreases as the distribution falls below threshold. Current efforts to more precisely measure the performance of the LAPPD with lower gains are in place by inserting an amplifier with the LAPPD.

\begin{figure} [!htbp]
\centering
\includegraphics[width=.5\linewidth]{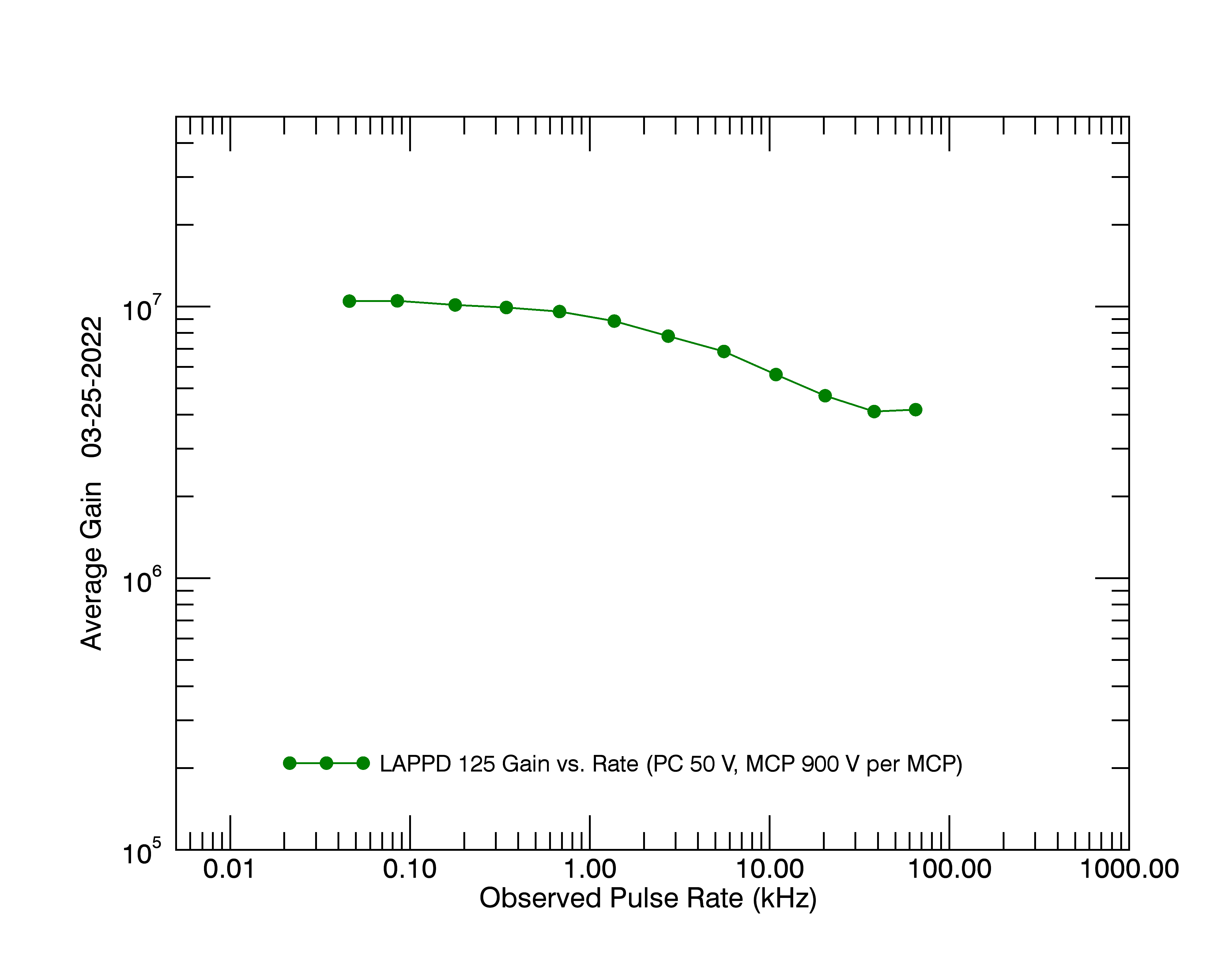}
\caption{LAPPD gain as a function of observed pulse rate. Gain is sustained at high rates, decreasing by a factor of 2 at 20 kHz/mm.}
\label{fig:gain:event}
\end{figure}

At high event rates, secondary electron emissions from the MCP channels weaken the internal electric field, decreasing incoming electron impact energy and subsequent secondary electron yields. The lost electrons are replenished via the MCP strip current, which is a function of the MCP resistance. A lower resistance MCP would be able to sustain its gain at higher event rates due to a higher MCP strip current at the cost of more power. The gain of LAPPD125 with 7 M$\Omega$ MCPs held up at high observed pulse rates, decreasing to half its nominal value at 20 kHz (Figure \ref{fig:gain:event}).

\subsection{Dark Count Rate}

\begin{figure}[htbp]
\centering 
\includegraphics[width=.4\textwidth]{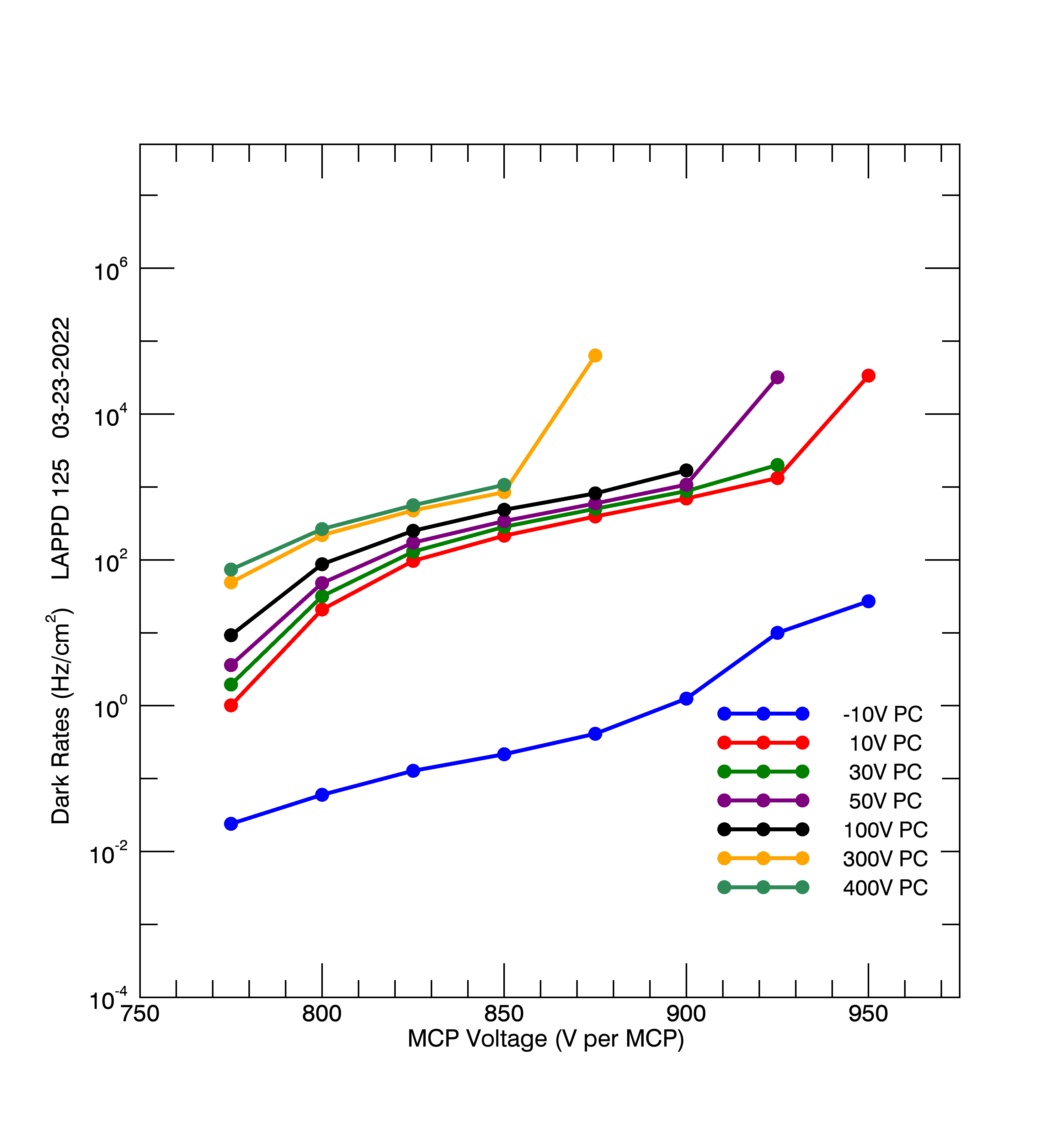}
\includegraphics[width=.4\textwidth]{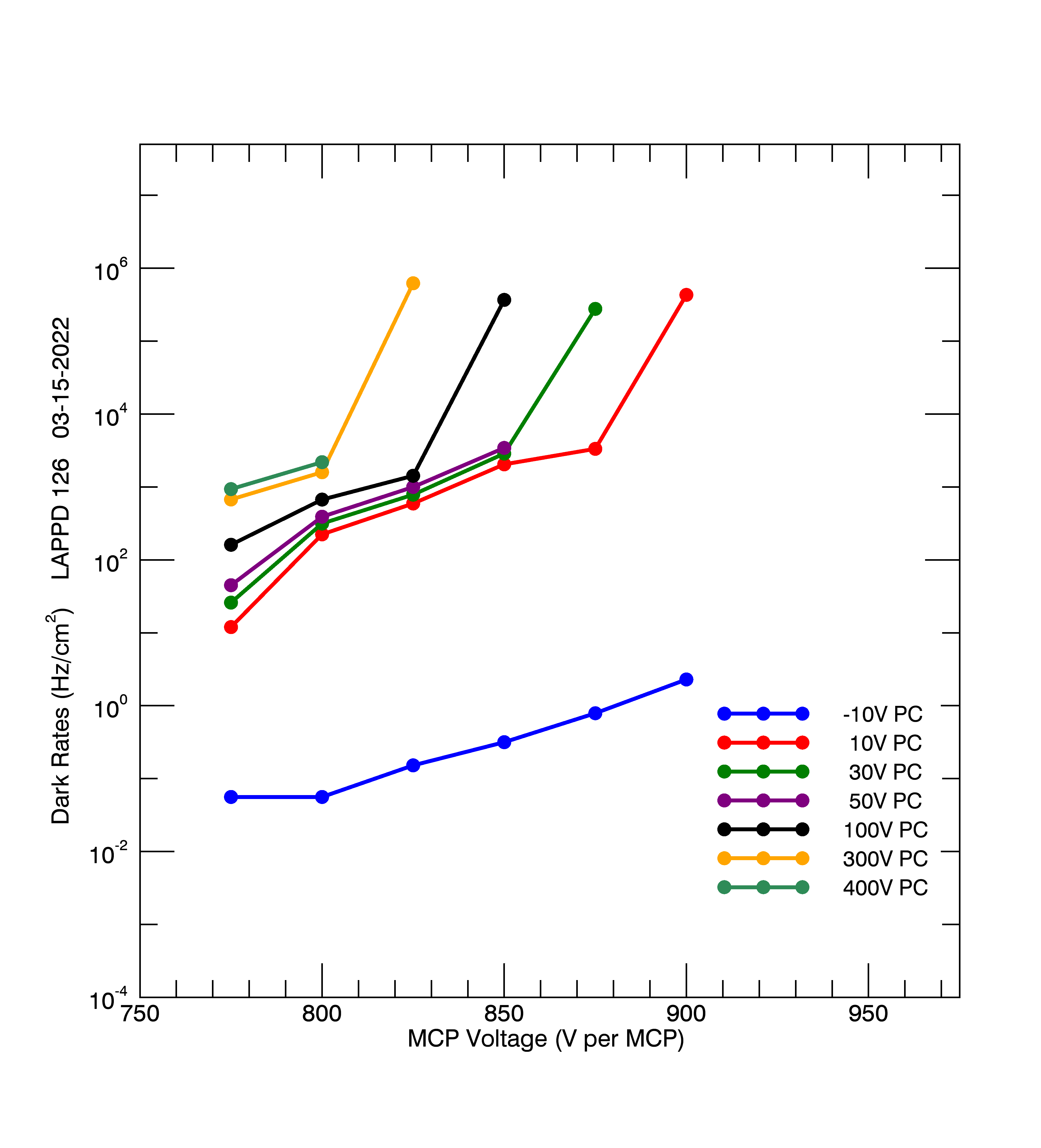}
\caption{\label{fig:dr:both} Dark rates as a function of both MCP voltage and photocathode voltage for LAPPD125 (left) and LAPPD126 (right). For most voltage configurations where the gain was $>10^6$, dark rates stay below 1 kHz/cm$^2$. "Spikes" in dark rates tend to decay over time.}
\end{figure}

When the photocathode voltage was positively biased relative to the entry MCP voltage, thermionic emission was suppressed and in this configuration, dark count rates as low as 0.03 Hz/cm$^2$ were recorded. In most applications, the photocathode voltage will be negatively biased. In either configurations, the dark count rates increased with both the MCP and photocathode voltages. The same principle that explains the increase in LAPPD gain applies to the increase in dark count rates. Free electrons generated within the LAPPD gain more energy, thus increasing secondary electron yields from the MCPs. At optimal voltage configurations, LAPPD125 and LAPPD126 exhibited high gains of $>$10$^7$ and $\sim3.5\cdot10^6$ respectively while sustaining relatively low dark rates around 1 kHz/cm$^2$.

Figure \ref{fig:dr:both} shows "spikes" in dark rates at certain voltage configurations. Dark rates at these states tend to decay over time if the LAPPD is left at these voltages. There is also a more rapid increase of the measured dark count rates at lower photocathode bias voltages due to the increase of gain and thus the increase of pulse discrimination efficiency. The measured and reported dark rates, which require transitions above and below the 4 mV threshold, actually correspond to much higher rates that are understated by the measurement method, typically on the order of MHz/cm$^2$.

\subsection{Timing Resolution}

\begin{figure}[htbp]
\centering 
\includegraphics[width=.45\textwidth]{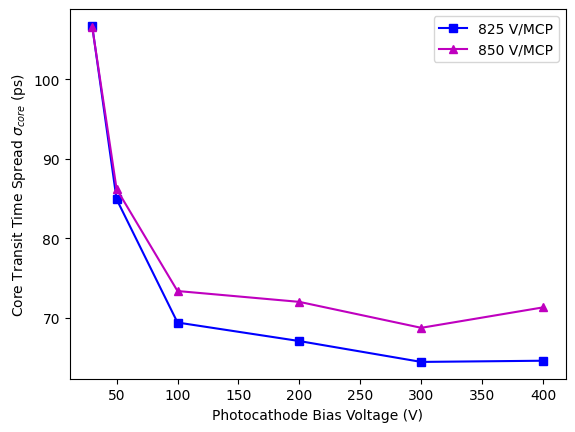}
\caption{\label{fig:tts:both} Core TTS $\sigma_\mathrm{core}$ measured as a function of the photocathode bias voltage for LAPPD125. Operation above 100 V yield the best timing results. The TTS is worse at 850 V/MCP than 825 V/MCP due to the consequence of increased gain and therefore increased small amplitude MCP dark pulses. This effectively creates a lower signal to noise ratio for TTS measurements.}
\end{figure}

$\sigma_\mathrm{core}$ improves significantly up to a 100 V bias between the photocathode and the entry MCP (Figure \ref{fig:tts:both}). PEs experience increased acceleration into the microchannels, therefore significantly reducing timing uncertainties to below 80 ps. Beyond 100 V, there is minimal improvement. At 825 V/MCP and a photocathode bias of 300 V, the LAPPD exhibited $\sigma_\mathrm{core}\approx$ 65 ps. As mentioned before, our results do not reflect the true timing capabilities of the LAPPD as the jitter in the DRS4 timesteps have not been accounted for. With improved calibrations of the readout board electronics, Vagnoni et al. \cite{LHCb} measured $\sigma_\mathrm{core}$ to be roughly 50 ps (SPE) and 8 ps for 20 PEs in a 15 mm radial spread.

\begin{figure}[htbp]
\centering 
\includegraphics[width=.45\textwidth]{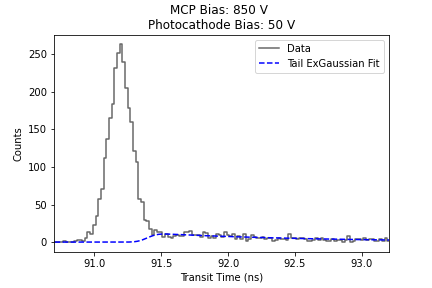}
\qquad
\includegraphics[width=.45\textwidth]{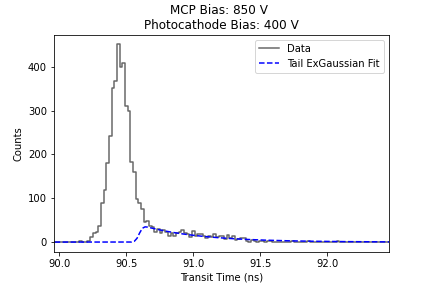}
\caption{\label{fig:frac:both} Transit time distributions for LAPPD125 at 850 V/MCP. At high photocathode bias voltages, the tail distribution becomes more prominent. This is an important point of consideration for photon counting applications which will be addressed in future efforts.}
\end{figure}

In photon counting applications, such as Cherenkov light detection in a water-based liquid scintillator neutrino detector \cite{cherenkov}, the tail distribution needs to be accounted for (Figure \ref{fig:frac:both}). Ref \cite{LHCb} found that the signal amplitudes originating from the tail distribution are lower than those from the core distribution. Thus, by setting a higher threshold on the incoming signal amplitude, the tail can be minimized. Therefore, the trade-off between higher photon collection efficiency and better timing resolution is dependent on the user and application needs. The overall timing performance can be improved by using smaller pore size MCPs. Detailed analyses on the LAPPD timing performance are currently in motion and will be reported in a separate publication.

\section{Capacitively Coupled LAPPD Position Resolution}

\begin{figure}[!htbp]
\centering 
\includegraphics[width=.4\textwidth]{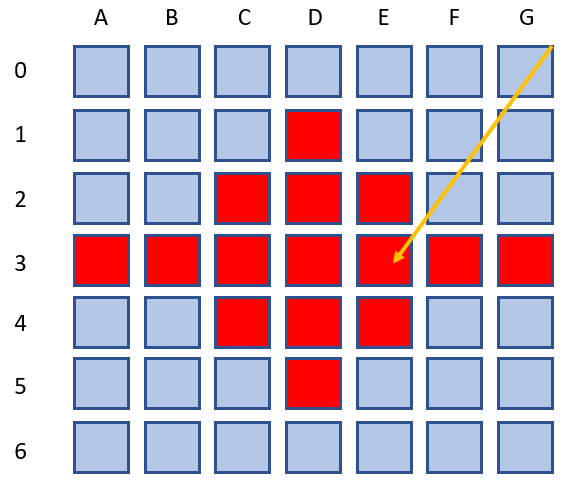}
\includegraphics[width=.5\textwidth]{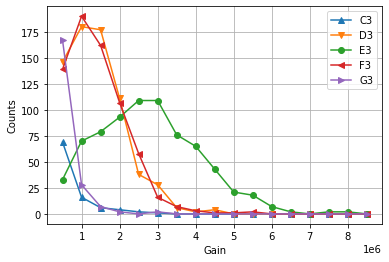}
\includegraphics[width=.45\textwidth]{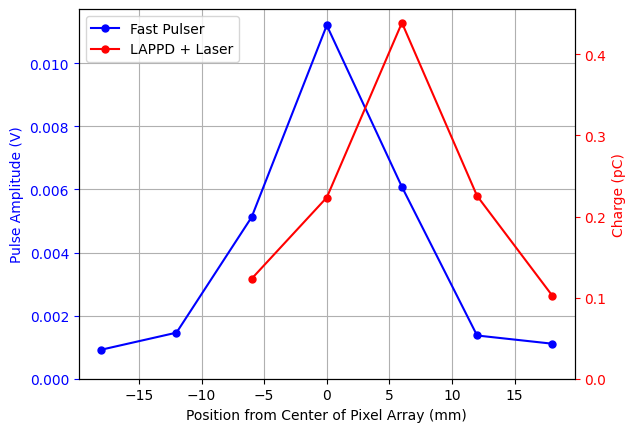}
\caption{\label{fig:latspread} A laser directed at the center of a 5 mm pixel at E3 (top left) was observed in adjacent pixels (top right). Red pixels denote those that were connected to a waveform sampler. The lateral amplitude profile was the same whether the pulse was applied as a voltage pulse at a point on a resistive anode, or by a laser on the window of a complete LAPPD (bottom). The pulses were observed in neighboring pixels at least 3 mm away from the point where the pulse was applied.}
\end{figure}

Position can be measured in the capacitively coupled Gen-II LAPPD by placing a patterned signal board beneath the anode. The interior side of the anode is coated with a resistive film. Electrons from the MCPs strike the resistive anode and diffuse to ground at the edges of the anode. The initial deposition on the anode is capacitively detected underneath as a fast-rising pulse.

This capacitively coupled technique allows flexibility in the signal board pattern, as there is no inherent pixelation anywhere in the anode. A signal board can be made with various size pixels, and even changed for another signal board as needed. A signal board with parallel striplines can also be manufactured, in which position along the stripline is determined by the relative time of pulse arrival at the ends of the stripline. This is an economical way to reduce electronics channels while covering a large area, although it does not easily handle simultaneous pulses on a single stripline.

The size of the pixels is related to the desired position resolution and to the width of the MCP pulse footprint on the anode. The width of the MCP pulse depends on the distance between the MCP and the anode, as well as the voltage between them. As the electrons make their way from the MCP to the anode, they will repel each other, thus widening their spatial distribution. If they propagate to the anode rapidly under a high voltage or a small gap, the spread will be minimized.

The lateral spread of the MCP charge pulse was discussed in Ref\cite{Saito}:

\begin{equation} \label{eq8}
     R = C \frac{Z}{\sqrt{V}},
\end{equation}
where
\begin{itemize}
    \item R = radius enclosing 70\% of the MCP charge cloud
    \item C = constant (2.7 to 3.4)
    \item Z = distance between MCP and anode (3-6 mm for the LAPPD)
    \item V = voltage between MCP and anode (typically 200 V in the LAPPD)
\end{itemize}
With the LAPPD, this radius may be calculated as $\sim$1.4 mm, where C = 3.3, Z = 6 mm, V = 200 V.

The potential at each pixel will depend on the distance of the pixel to the MCP charge. Consequently, the MCP charge may be detected at lateral distances greater than $\sim$1.4 mm. For example, a simulation was performed by applying a fast voltage pulse with a risetime similar to an MCP pulse, 0.7 ns, to the resistive anode of an LAPPD tile base. The pulse was capacitively coupled to a signal board with various size pixel widths, and the amplitude was measured on neighboring pixels. An example is shown in Figure \ref{fig:latspread} for 5 mm pixels, with a 6 mm pitch. The pulse was observed at approximately half amplitude in the two neighboring pixels. The measurement was repeated with a laser, using a spot illumination of $\sim$1mm diameter, and the lateral spread was the same (bottom, Figure \ref{fig:latspread}). The charge deposition was therefore observable at $\sim$3mm to either side of the point of deposition.

\begin{figure} [!htbp]
\centering
\includegraphics[width=.5\linewidth]{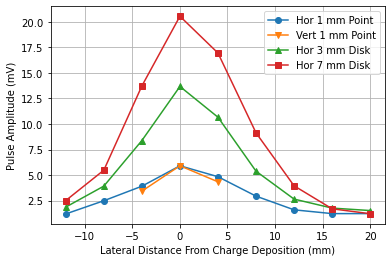}
\caption{The point current in Figure \ref{fig:latspread} was applied to the resistive anode and observed capacitively with 3 mm pixels. A similar result was obtained --- the two neighboring pixels at $\pm4$ mm  observed the signal at approximately more than half amplitude. The next neighbors also observed the pulse, but at a diminished amplitude, approximately 8 mm away. The same result was obtained when the current was applied to the resistive anode with an aluminum disk, representing a broader pulse footprint on the anode.}
\label{fig:3mm}
\end{figure}

A similar result was obtained with 3 mm pixels (Figure \ref{fig:3mm}). The pulse applied to the center pixel via a 1 mm diameter spring tip pin to the resistive anode was observed in the two neighbors, and at a diminished amplitude in the two neighbors beyond them, for a total of $\sim$8 mm to either side of the center. Interestingly, a similar result was obtained when the pulse was applied to the resistive anode via a spring tip using a 1.5 mm thick disk of 3 or 7 mm in diameter, representing a broader MCP pulse footprint. The lateral response to the pulse was $\sim$12 mm FWHM.

\begin{figure} [!htbp]
\centering
\includegraphics[width=.45\linewidth]{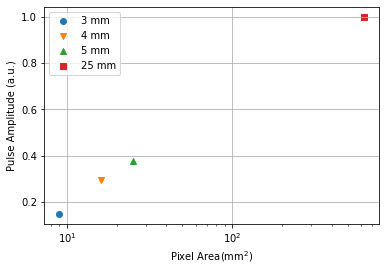}
\caption{The relative pulse height is shown for various size pixels. The smaller pixels represent a partial detection of the signal, as the effective footprint of the MCP pulse is wider than 3-4 mm.}
\label{fig:nalu}
\end{figure}

The size of the pixel is related to the measured amplitude. In part, this is due to the cutoff by the capacitance between the pixel and the anode. Moreover, the effective footprint of the MCP pulse is wider than the smallest pixels, so a single pixel represents a partial detection of the MCP pulse. The relative pulse height on single small pixels using a laser is compared to the response from a 25 mm square pixel in Figure \ref{fig:nalu}.

\begin{figure}[!htbp]
\centering 
\includegraphics[width=.45\textwidth]{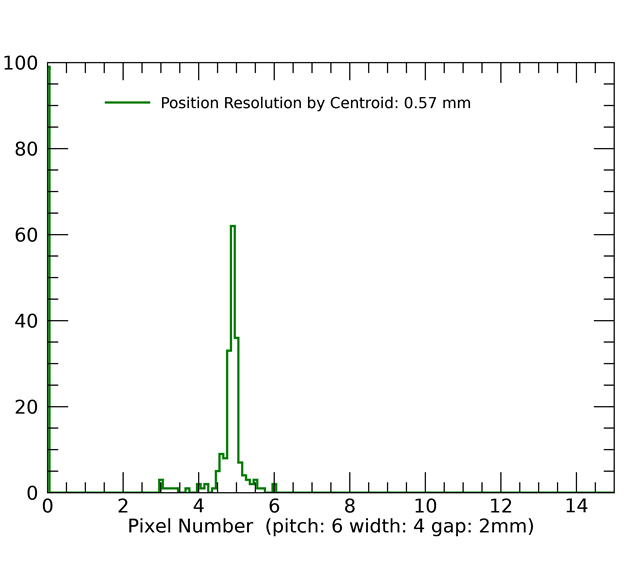}
\quad
\includegraphics[width=.45\textwidth]{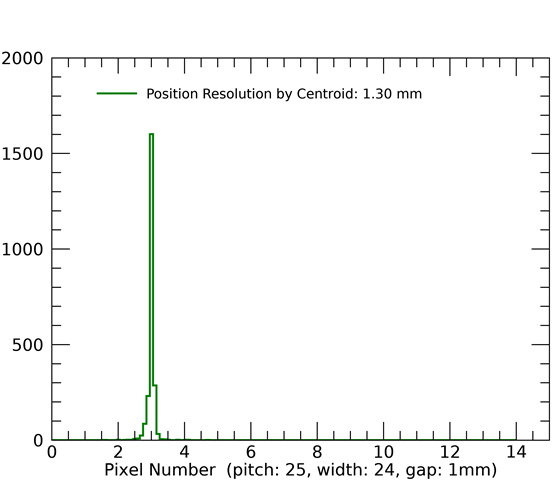}
\caption{\label{fig:centroid} Position as calculated by centroiding nearest neighbor pixels is shown for 6 mm (left) and 25 mm (right) pitch pixels.}
\end{figure}

The position may be calculated by centroiding the amplitudes of neighboring pixels. This improves the resolution over the pixel width itself. Position resolution is calculated for 6 and 25 mm pitch pixel sizes and as shown in Figure \ref{fig:centroid}, is 0.57 mm and 1.30 mm respectively with centroiding. The pulses were responses to a laser on an assembled LAPPD.

\section{Conclusions}

\begin{table}[ht]
    \caption{Capacitively Coupled LAPPD Performance Results Summary} 
    \centering 
    \setlength{\extrarowheight}{.5em}
    \begin{tabular}{c| c c} 
        \hline\hline 
        Parameters & Typical Specifications & Recent Specifications  \\ [0.5ex] 
        \hline 
        Gain & $>5\cdot10^6$ & up to 10$^7$  \\ 
        QE & $>$20\% at 365 nm & $>$30\% at 365 nm  \\
        Dark Count Rates & $<$10 kHz/cm$^2$ & $\sim$1 kHz/cm$^2$  \\
        Single PE Timing Resolution & $<$100 ps & $\sim$65 ps \\
        Position Resolution & \multicolumn{2}{c}{$\mathcal{O}$(mm) (dependent on readout board)}  \\ [1ex] 
        \hline 
    \end{tabular}
    \label{table:nonlin} 
\end{table}

The Gen-II LAPPD with the capacitively coupled readout has demonstrated performance results similar to those of the Gen-I LAPPD with the stripline direct readout, as shown in Table \ref{table:nonlin}. The former offers the advantage of user-friendly flexibility in designing the external readout board, allowing the Gen-II LAPPD to be both multi-purpose and application-specific.

The LAPPD is an excellent candidate for collider experiments for HEP and NP research, non-accelerator neutrino and rare-event experiments, TOF-PET systems, and nuclear non-proliferation applications. The LAPPD is currently in full production mode at Incom, Inc. Both Gen-I and Gen-II LAPPDs, glass or ceramic, are available today with either 10 $\mu$m or 20 $\mu$m pore size MCPs upon request.


\acknowledgments

This work was supported by the U.S. Department of Energy, USA, Offices of
High Energy Physics, USA and Nuclear Physics, USA under DOE contracts: DOE DE-SC0019821, DOE DE-SC0018778, DOE DE-SC0017929, DOE DE-SC0021782, DOE DE-SC0020578.




\begin{thebibliography}{99}



\bibitem{EIC}
R. Abdul Khalek, U. D'Alesio, M. Arratia, et al., \emph{Snowmass 2021 White Paper: Electron Ion Collider for High Energy Physics}, \href{https://arxiv.org/abs/2203.13199}{arXiv:2203.13199}.

\bibitem{RICH}
A. Kiselev, et al.,
\emph{Capacitively Coupled LAPPDs with 2D Pixelated Readout Planes for Time of Flight and Ring Imaging Cherenkov Applications}, 
\href{https://www.osti.gov/biblio/1854097}{\emph{IEEE Nuclear Science Symposium (NSS) and Medical Imaging Conference (MIC)} (2021)}.

\bibitem{LHCb}
Stefano Perazzini, Fabio Ferrari, Vincenzo Maria Vagnoni
and on behalf of the LHCb ECAL Upgrade-2 R\&D Group, \emph{Development of an MCP-Based Timing Layer for the LHCb
ECAL Upgrade-2},
\href{https://doi.org/10.3390/instruments6010007}{\emph{Instruments {\bf 2022}, 6, 7}}.

\bibitem{DUNE}
The DUNE Collaboration, \emph{Deep Underground Neutrino Experiment (DUNE) Near Detector Conceptual Design Report}, \href{https://doi.org/10.3390/instruments5040031}{\emph{Instruments {\bf 2021}, 5, 31}}.

\bibitem{ANNIE}
The ANNIE Collaboration, \emph{Accelerator Neutrino Neutron Interaction Experiment (ANNIE): Preliminary Results and Physics Phase Proposal}, \href{https://arxiv.org/abs/1707.08222}{arXiv:1707.08222}.

\bibitem{WATCHMAN}
C. Grant and on behalf of the AIT-WATCHMAN Collaboration,
\emph{WATCHMAN: A Remote Reactor Monitor and Advanced Instrumentation Testbed},
\href{https://iopscience.iop.org/article/10.1088/1742-6596/1468/1/012182}{2020 \emph{J. Phys.: Conf. Ser.} {\bf 1468}}.

\bibitem{THEIA}
M. Askins, et al., \emph{THEIA: an advanced optical neutrino detector}, \href{https://doi.org/10.1140/epjc/s10052-020-7977-8}{\emph{Eur. Phys. J. C} {\bf 80}, 416 (2020)}.

\bibitem{TOFPET}
W. Worstell, et al.,
\emph{Characterization of coincident single photoelectron time response of a Large Area Picosecond Photon Detector (LAPPD) for Time-of-Flight PET Application},
\href{https://jnm.snmjournals.org/content/62/supplement_1/1131}{\emph{Journal of Nuclear Medicine May 2021}, {\bf 62} (supplement 1) 1131 (2021)}.

\bibitem{SIPM}
SiPM, Hamamatsu Photonics, K. K.,
\href{https://www.hamamatsu.com/us/en/product/optical-sensors/mppc.html}{https://www.hamamatsu.com/us/en/product/optical-sensors/mppc.html}.

\bibitem{hamamatsu}
Hamamatsu Photonics, K. K.,
\emph{PHOTOMULTIPLIER TUBES - Basics and Applications (Edition 3a)}, (2007)
\url{https://www.hamamatsu.com/content/dam/hamamatsu-photonics/sites/documents/99_SALES_LIBRARY/etd/PMT_handbook_v3aE.pdf}.

\bibitem{MPPC}
MPPC, Hamamatsu Photonics, K. K.,
\href{https://www.hamamatsu.com/content/dam/hamamatsu-photonics/sites/documents/99_SALES_LIBRARY/ssd/s13360_series_kapd1052e.pdf}{https://www.hamamatsu.com/content/dam/hamamatsu-photonics/sites/documents/99\_SALES\_LIBRARY/ssd/s13360\_series\_kapd1052e.pdf}.

\bibitem{PLANACON}
Planacon, Photonis,
\href{https://www.photonis.com/products/planacon}{https://www.photonis.com/products/planacon}.

\bibitem{POPECKI}
M. A. Popecki, et al.,
\emph{Microchannel plate fabrication using glass capillary arrays with Atomic Layer Deposition films for resistance and gain},
\href{https://doi.org/10.1002/2016JA022580}{\emph{J. Geophys. Res. Space Physics}, 121, 7449-7460 (2016)}.

\bibitem{LYASHENKO2020}
A. Lyashenko, et al.,
\emph{Performance of Large Area Picosecond Photo-Detectors (LAPPD\textsuperscript{TM})},
\href{https://doi.org/10.1016/j.nima.2019.162834}{\emph{Nucl. Instr. and Meth. A} 958 (2020) 162834}
[\href{https://arxiv.org/abs/1909.10399}{arXiv:1909.10399}].


\bibitem{SPE}
J. Wang, et al.,
\emph{Development and testing of cost-effective, 6 cm × 6 cm
MCP-based photodetectors for fast timing applications},
\href{https://doi.org/10.1016/j.nima.2015.09.020}{\emph{Nucl. Instr. and Meth. A} 804 (2015) 84} [\href{https://arxiv.org/abs/1604.07738}{arXiv:1604.07738}].

\bibitem{DRS4}
S. A. Ritt,
\emph{Design and performance of the 6 GHz waveform digitizing chip DRS4},
\href{https://doi.org/10.1109/NSSMIC.2008.4774700}{\emph{2008 IEEE Nuclear Science Symposium conference Record} (2008) 1512--1515}.

\bibitem{Deadtime}
A.S. Tremsin, J.F. Pearson, G.W. Fraser, W.B. Feller, P. White,
\emph{Microchannel plate operation at high count rates: new results},
\href{https://doi.org/10.1016/0168-9002(96)00482-2}{\emph{Nucl. Instr. and Meth. A} 379 (1996) 1}.

\bibitem{calibration}
H. Kim et al.,
\emph{A new time calibration method for switched-capacitor array-based waveform samplers},
\href{https://doi.org/10.1016/j.nima.2014.08.025}{\emph{Nucl. Instr. and Meth. A} 767 (2014) 67}.

\bibitem{UQE}
K. Nakamura, Y. Hamana, Y. Ishigami, and T. Matsui,
\emph{Latest bialkali photocathode with ultra high sensitivity},
\href{https://doi.org/10.1016/j.nima.2010.02.220}{\emph{Nucl. Instr. and Meth. A} 623 (2010) 1}.

\bibitem{cherenkov}
Kaptanoglu, T., Callaghan, E.J., Yeh, M. , et al.,
\emph{Cherenkov and scintillation separation in water-based liquid scintillator using an LAPPD \textsuperscript{TM}}, 
\href{https://doi.org/10.1140/epjc/s10052-022-10087-5}{\emph{Eur. Phys. J. C} {\bf 82,} 169 (2022)}
[\href{https://arxiv.org/abs/2110.13222}{arXiv:2110.13222}].

\bibitem{Saito}
M. Saito, Y. Saito, K. Asamura, T. Mukai,
\emph{Spatial charge cloud size of microchannel plates},
\href{https://doi.org/10.1063/1.2472595}{\emph{Review of Scientific Instruments} {\bf 78}, 023302 (2007)}.

\bibitem{Rossi}
L. Rossi, et al.,
\emph{Pixel Detectors: From Fundamentals to Applications},
\href{https://link.springer.com/book/10.1007/3-540-28333-1}{\emph{Springer Science \& Business Media} (2006)}.






\end{thebibliography}
\end{document}